\documentclass[12pt,semidraft]{article}

\usepackage{array} 
\usepackage{epsfig}
\usepackage{amssymb}
\usepackage{amsmath}                          
\usepackage{graphics,graphpap}
\usepackage{graphicx}
\usepackage{epstopdf}

\renewcommand{\thefootnote}{\fnsymbol{footnote}}

\newcommand{\bea}{\begin{eqnarray}}
\newcommand{\eea}{\end{eqnarray}}
\newcommand{\bac}{\begin{array}{c}}
\newcommand{\bacc}{\begin{array}{cc}}
\newcommand{\baccc}{\begin{array}{ccc}}
\newcommand{\ea}{\end{array}}
\newcommand{\bc}{\begin{center}}
\newcommand{\ec}{\end{center}}
\newcommand{\bt}{\begin{tabular}}
\newcommand{\et}{\end{tabular}}
\newcommand{\bit}{\begin{itemize}}
\newcommand{\eit}{\end{itemize}}

\newcommand{\done}{{\rm d}}

\newcommand{\DO}{D$0\!\!\!/$ }
\newcommand{\nnb}{\nonumber}

\def\fslash#1{\setbox0=\hbox{$#1$}%
\rlap{\ifdim\wd0>.7em\kern.22\wd0\else\kern.1\wd0\fi /}#1}

\def\simge{\mathrel{%
   \rlap{\raise 0.511ex \hbox{$>$}}{\lower 0.511ex \hbox{$\sim$}}}}

\def\simle{\mathrel{
   \rlap{\raise 0.511ex \hbox{$<$}}{\lower 0.511ex \hbox{$\sim$}}}}

\newcommand{\mph}[1]{ m_{#1,{\rm phys}}  }
\newcommand{\phys}[2]{ {#1}_{#2,{\rm phys}} }
\newcommand{\physs}[1]{{#1}_{\rm phys} }

\begin{document}


\begin{titlepage}
\begin{flushright}\begin{tabular}{l}
IPPP/07/49\\
DCPT/07/98
\end{tabular}
\end{flushright}
\vskip1.5cm
\begin{center}
   {\Large \bf \boldmath The process $gg \to h_0 \to \gamma\gamma$  
   \\[0.3cm]
   in the Lee-Wick Standard Model}
    \vskip1.3cm {\sc
 F.~Krauss\footnote{frank.krauss@durham.ac.uk},
 T.~E.~J.~Underwood\footnote{t.e.j.underwood@durham.ac.uk}\footnote{New address:
   {\em Max-Planck-Institut f\"ur Kernphysik, Saupfercheckweg 1, 69117
     Heidelberg, Germany}}, and
 R.~Zwicky\footnote{roman.zwicky@durham.ac.uk}
   \vskip0.5cm
        {\em IPPP, Department of Physics, 
Durham University, Durham DH1 3LE, UK}} \\
\vskip1.5cm 
\vskip0.5cm

{\large\bf Abstract:\\[8pt]} \parbox[t]{\textwidth}{ The process $gg \to h_0
  \to \gamma \gamma$ is studied in the Lee-Wick extension of the Standard
  Model (LWSM) proposed by Grinstein, O'Connell and Wise.  In this model
  negative norm partners for each SM field are introduced with the aim to
  cancel quadratic divergences in the Higgs mass. All sectors of the model
  relevant to $gg \to h_0 \to \gamma \gamma$ are diagonalized and results are
  commented on from the perspective of both the Lee-Wick and higher derivative
  formalisms. Deviations from the SM rate for $gg\to h_0$ are found to be of
  the order of 15\% -- 5\% for Lee-Wick masses in the range 500~GeV --
  1000~GeV.  Effects on the rate for $h_0\to \gamma \gamma$ are smaller, of
  the order of 3\% -- 1\% for Lee-Wick masses in the same range.  These
  comparatively small changes may well provide a means of distinguishing the
  LWSM from other models such as universal extra dimensions where same-spin
  partners to Standard Model fields also appear. Corrections to determinations
  of CKM elements $|V_{t(b,s,d)}|$ are also considered and are shown to be
  positive, allowing the possibility of measuring a CKM element larger than
  unity, a characteristic signature of the ghost-like nature of the Lee-Wick
  fields.}

\vfill

\end{center}
\end{titlepage}

\setcounter{footnote}{0}
\renewcommand{\thefootnote}{\arabic{footnote}}

\newpage

\section{Introduction}

The Higgs mechanism is one of the cornerstones of the Standard Model (SM) since 
it provides an elegant way of endowing the fundamental particles of the SM with 
mass.  Therefore, it is not surprising that in past decades large efforts have 
been undertaken to find the Higgs boson and thus confirm this mechanism of
electroweak symmetry breaking (EWSB).  From a more theoretical point of view,
the Higgs mechanism, however elegant, has a severe aesthetical flaw on the
quantum level.  Due to the emergence of quadratic divergences in the
self-energy corrections of the scalar Higgs boson, there must be some
essential fine-tuning in order for its physical mass to satisfy all
constraints, including the upper limit of roughly 1 TeV stemming from
unitarity requirements in $W_LW_L$ scattering.  One viable method to
significantly reduce the amount of fine-tuning consists of introducing new
degrees of freedom such that the quadratic divergences disappear.  To protect
the disappearance, additional symmetries, like in the case of supersymmetry,
have been postulated, or the mechanism of EWSB and the Higgs boson itself have
been explained on different dynamical grounds like in the case of technicolour
models, which render the Higgs boson a composite object.  Furthermore, in the
past years a number of models involving extra dimensions have been developed
and discussed which aim to minimise the
amount of fine tuning by reducing the upper scale of the theory.   

On the basis of the work ``Finite Theory of QED'' by Lee and Wick (LW)
\cite{negmetric,finiteqed}, Grinstein, O'Connell and Wise (GOW) \cite{GOW}
proposed an extension of the SM which is free from quadratic divergences
\footnote{ However, it is worth emphasizing that the Lee-Wick SM is not
  finite, contrary to LW QED \cite{GOW}.}. The basic idea of Lee and Wick was
to assume that the regulator employed in the framework of Pauli-Villars
regularisation indeed is a physical degree of freedom which is not removed
from the amplitudes. The scale of the LW masses has to be high enough to evade
current experimental constraints and low enough in order to solve the
hierarchy problem.  Moreover, in the context of the see-saw mechanism for
neutrino masses, it was shown in Ref.~\cite{nuR_hierarchy} that the presence
of a very heavy right handed neutrino does not introduce destabilizing
corrections to the Higgs mass of the form $\delta m_H^2 \sim m_{\nu_R}^2$,
provided that $m_{\nu_R} \gg m_{\nu_L}$.  This is unlike the minimal see-saw
extension of the SM, where destabilizing corrections of the type mentioned
above do appear.

However, a number of questions emerge concerning the interpretation and
consequences of the Pauli-Villars ``wrong-sign'' states.  A formal argument in
favour of unitarity was given in Refs. \cite{finiteqed,Lee-Erice}, based on
the idea that once interactions are switched on, the negative norm states mix
with physical multiparticle states into states with complex masses. The
orthogonality of those states with the physical states then ensures that the
S-matrix does map one type into the other. This is a necessary condition for
the unitarity of the S-matrix restricted to the physical (positive norm)
subspace.  An illustrative example of how the finite width preserves unitarity
can be found in \cite{GOW,Coleman-Erice}.  Contrary to the regular finite
width, the complex masses are situated on the physical sheet upsetting the
usual analytic structure. This demands a modification to the integration
contour in the complex plane in order to eliminate exponentially growing modes
\cite{finiteqed,S-gang}.  Consistent results using this approach were obtained
order-by-order in perturbation theory in Refs. \cite{Lee-Erice,Coleman-Erice}.

This modification of the integration contour leads to acausal effects
\cite{finiteqed,Lee-Erice,Coleman-Erice} on the time scale of the inverse
width of the LW state.  The width $\Gamma_{\rm LW}$ of the LW resonances under
consideration is too large for the time scales which can be resolved at
current colliders. On the other hand, in this paper, finite width effects will
be neglected since the LW mass scales and the difference between the LW and SM
masses are much larger than the LW width, $m_{\rm LW} \gg \Gamma_{\rm LW}$ and
$(m_{\rm LW}-m_{\rm SM}) \gg \Gamma_{\rm LW}$.

A non-perturbative definition of the Lee-Wick approach using path integrals
was first investigated in Ref.~\cite{BoulawareGross} but, as the authors
state, the relation to the LW prescription remained unclear.  The Euclidian
path integral in the Higgs sector was then studied almost ten years later
\cite{Jansen_formal,Jansen_lattice}, with the aim of investigating the
triviality bound on the Higgs mass from the lattice using a symmetry
preserving regularization. There it was clearly stated that the ghost poles
due not allow a continuation to Minkowski space. Recently a path integral
formulation has been advocated based on a test function space \cite{tonder},
which allows the continuation to a convergent Minkowski space formulation.
Moreover, in that work the Lee-Wick contour prescription was derived from a
path integral approach. It is worth mentioning that this prescription does not
correspond to adding convergence factors to the action and therefore clarifies
why earlier attempts were not able to relate the Euclidian path integral and
the LW formulation.

In their recent paper, GOW made use of the fact that regularising
Pauli-Villars-like states may be obtained from higher derivative terms in the
Lagrangian.  To illustrate this and to show how the additional fields of the LW
approach cancel the quadratic divergences in the LW model, a toy model will be
considered, similar to the one proposed by GOW \cite{GOW}, of a
self-interacting, real scalar field $\hat\phi$, with a higher derivative term
in the Lagrangian density
\bea
{\cal L}_{\rm hd} =
\frac12(\partial_\mu\hat\phi)(\partial^\mu\hat\phi) - \frac12m^2\hat\phi^2 -
\frac{1}{2M^2}(\partial^2\hat\phi)^2 - \frac{\lambda}{4!}\hat\phi^4\,.
\label{eq:hdphi}
\eea
Taken directly from this Lagrangian, the propagator of the $\hat\phi$ field
reads \bea
\label{eq:prop}
\hat D(p) = \frac{i}{p^2 - p^4/M^2 - m^2} = \frac{-i M^2}{(p^2 - \physs{m}^2)(p^2 - \physs{M}^2)}\,,
\eea
in momentum space, where 
\begin{equation}
\label{eq:mMphys}
\physs{m}^2[\physs{M}^2] = \frac{1}{2} \left(M^2 -[+] \sqrt{M^4 - 4 m^2 M^2}\right)\,,
\end{equation}
are the physical masses.  A possible interpretation is that, effectively, this
theory describes two degrees of freedom with masses $\physs{m}$ and
$\physs{M}$, respectively.  The authors of \cite{GOW} have shown how to make
these degrees of freedom manifest by introducing an auxiliary ``Lee-Wick''
field $\tilde\phi$ of mass $M$, such that the Lagrangian becomes
\bea
{\cal L}_{\rm aux} =
\frac12(\partial_\mu\hat\phi)(\partial^\mu\hat\phi) -
\frac12m^2\hat\phi^2 -
\tilde\phi\partial^2\hat\phi +
\frac{1}{2}M^2\tilde\phi^2 -
\frac{\lambda}{4!}\hat\phi^4\,.
\label{eq:auxphi}
\eea
The higher derivative Lagrangian Eq.~\eqref{eq:hdphi} may be recovered from
Eq. \eqref{eq:auxphi} by employing the equation of motion (EoM)
$M^2\tilde\phi= \partial^2\hat\phi$. Furthermore, by decomposing the field of
the higher-derivative theory $\hat\phi$ into a ``standard'' field $\phi$ and a
LW field $\tilde\phi$, through the shift $\hat\phi=\phi-\tilde\phi$, the
Lagrangian can be written
\bea {\cal L}_{\rm eff} =
\frac12(\partial_\mu\phi)(\partial^\mu\phi) -
\frac12(\partial_\mu\tilde\phi)(\partial^\mu\tilde\phi) -
\frac12m^2(\phi-\tilde\phi)^2 + \frac{1}{2}M^2\tilde\phi^2 -
\frac{\lambda}{4!}(\phi-\tilde\phi)^4\,.  \eea
Clearly, in the presence of $m$, the two states $\phi$
and $\tilde\phi$ mix. The Lagrangian can be expressed in terms of mass
eigenstates after a symplectic transformation on the fields
\bea
\left(\bac\phi\\\tilde\phi\ea\right) =
\left(\bacc\cosh\theta & \sinh\theta\\
  \sinh\theta & \cosh\theta\ea\right)
\left(\bac\phi'\\\tilde\phi'\ea\right)\,, \qquad \tanh 2\theta =
\frac{-2m^2/M^2}{1-2m^2/M^2}\,, \eea where the constraint, $2\,m < M$ must be
satisfied, leading to the constraint on the physical masses $\physs{m} <
\physs{M} $. Expressed in the new basis, the Lagrangian density
becomes \bea\label{Eq:EffToyLagrangian} {\cal L}_{\rm eff} =
\frac12(\partial_\mu\phi')(\partial^\mu\phi') -
\frac12(\partial_\mu\tilde\phi')(\partial^\mu\tilde\phi') -
\frac12{\physs{m}}^2\hat{\phi'}^2 + \frac12{\physs{M}}^2\tilde{\phi'}^2 -
\frac{\physs{\lambda}}{4!}(\phi-\tilde\phi)^4\,.  \eea The ultraviolet
behaviour of this theory can be accessed from the Lagrangian of Eq.\
(\ref{Eq:EffToyLagrangian}), where from now on the primes are dropped.  At
order ${\cal O}(\lambda)$, for instance, the self-energy correction to the
$\phi$ field has two tadpole contributions, where either $\phi$ or
$\tilde\phi$ run in the loop.  Employing dimensional regularisation with
$d=4-2\varepsilon$, they read
\begin{equation*}
\Sigma(p^2)
=
i\lambda\Big( \int\frac{\done^dk}{(2\pi)^d}\frac{i}{k^2-m^2} -
\int\frac{\done^dk}{(2\pi)^d}\frac{i}{k^2-M^2}\Big) =
i\lambda\int\frac{\done^dk}{(2\pi)^d}
            \frac{i(m^2-M^2)}{(k^2-m^2)(k^2-M^2)}\,,
\end{equation*}
and apparently the quadratic divergence has been reduced to a logarithmic one.
This is the anticipated result, since the introduction of higher-derivative
terms in the way proposed by GOW leads to higher powers of the momentum in the
propagators of the full theory, thus improving the convergence of graphs.
Effects of higher derivative terms in the interactions could eventually upset
this picture, however such terms are not present in the toy model discussed
here.  For the full LW version of the SM, studied in Sec.\ \ref{sec:LW-SM},
the situation is not so clear.

In the past few months various consequences of the LW extension of the SM
(LWSM) have been discussed, in particular the prospects for finding LW bosons
at the LHC \cite{Rizzo:2007ae} and LW effects in flavour physics
\cite{Dulaney:2007dx}.  This publication aims at a first study of the
phenomenology of the LWSM at loop level in the ``golden-plated'' discovery
modes for low-mass Higgs bosons at the LHC, $gg\to h_0\to\gamma\gamma$.
Following the reasoning above, one would naively expect similarly strong
cancellations in all kinds of loop-induced processes, effectively suppressing
a plethora of interesting signals, such as, e.g.\ $gg\to h_0\to\gamma\gamma$.

So the outline of this paper is as follows; in section \ref{sec:LW-SM} the
initial work of GOW on the construction of the LWSM will be supplemented with
a discussion of all transformations from the original degrees of freedom into
the mass eigenstates.  Then the exact cancellation of quadratic divergences in
the Higgs self-energy will be checked. In section \ref{sec:gluephoton}, the
decay widths for $h_0\to gg$ and $h_0\to\gamma\gamma$ will be calculated in
the LWSM.  They will be multiplied such that the overall cross section for
$gg\to h_0\to\gamma\gamma$ can be obtained and compared to results from other
models, such as universal extra dimensions (UEDs).  Furthermore, in this
section, some of the phenomenological consequences will be highlighted.
Section \ref{sec:flavour} will outline some interesting effects in the flavour
physics sector, in particular related to the determination of CKM matrix
elements. Some short discussion of potential further effects of this model and
how it can be distinguished from other models beyond the Standard Model round
off the publication, before its central findings are summarised.

\section{The Lagrangian of the Lee-Wick Standard Model}
\label{sec:LW-SM}

\subsection{Construction principle}

The principle idea underlying the Lee-Wick extension of the SM constructed by
GOW is to augment the SM Lagrangian with higher derivative terms. Of course,
in principle a huge number of such terms are allowed, see for instance
\cite{Antoniadis:2007xc,Ghilencea:2007ex}, but only few of them are actually
selected for the construction of the LWSM. In particular; \bit
\item for each scalar, a term
	\bea\label{eq:HDscalar}
	-\frac{1}{M_\phi^2}\,
	\big(\hat{D}_\mu \hat{D}^\mu \hat{\phi} \big)^\dagger 
	\big(\hat{D}_\nu \hat{D}^\nu \hat{\phi} \big)
	\eea
	is introduced (see also the toy model in the introduction), 
	where $D^\mu$ is the gauge covariant derivative;
\item for each fermion, a term
	\bea\label{eq:HDfermion}
	+\frac{1}{M_\psi^2}\,
	\overline{\hat{\psi}}\,i \hat{\fslash{D}} \hat{\fslash{D}}
	\hat{\fslash{D}}\,\hat{\psi}
	\eea
	is added;
\item and for each gauge field, a term
	\bea\label{eq:HDvector}
	+\frac{1}{M_A^2}\,{\rm Tr} 
	\big(\hat{D}^\mu \hat{F}_{\mu \nu}\big)\,
	\big(\hat{D}^\lambda \hat{F}_\lambda{}^\nu \big)
	\eea
	is added.
The field strength tensor is generalised to \bea \hat{F}^{\mu\nu}=
	\partial_\mu \hat{A}_\nu - \partial_\nu \hat{A}_\mu
	-i g [\hat{A}_\mu, \hat{A}_\nu]
	\;\;\;\mbox{\rm with}\;\;\;
	\hat{A}_\mu = \hat{A}_\mu^A\,T^A\,,
	\eea
	where $T^A$ are the generators of the corresponding gauge 
	group with coupling constant $g$.\eit

It should be stressed that the rest of the structure of the SM, in particular
the interactions, remains unchanged apart from replacing the original SM
fields with the ``hatted'' fields of the higher derivative theory.

\subsection{Higgs sector}
\label{Sec:HiggsBosons}

In the Higgs sector of the LWSM, higher derivative terms of the form of Eq.\
\eqref{eq:HDscalar} are eliminated in a way identical to the toy model;
auxiliary fields $\tilde{H}$ are introduced and their EoM are used to replace
the higher derivative terms with mass terms and kinetic terms for the
auxiliary fields.  The kinetic terms between $\hat{H}$ and $\tilde{H}$ are
diagonalized by applying a shift $\hat{H} = H - \tilde{H}$.  Further
diagonalization of the Higgs mass matrix proceeds in the same fashion as in
the toy model studied by GOW and in the introduction. Thus, in the unitary
gauge the two doublets are
\begin{equation}
\label{eq:unitarygauge}
H^\top = \big[0, (v+h_0)/\sqrt{2}\big]\,,\quad 
\tilde{H}^\top = 
\big[\tilde{h}_+,(\tilde{h}_0+i\tilde{P}_0)/\sqrt{2}\big]\,,
\end{equation} 
where $h_0$ refers to the SM-like neutral Higgs boson, $\tilde h_+$,
$\tilde h_0$, and $\tilde P_0$ refer to the charged and neutral 
scalar and pseudoscalar LW Higgs bosons, respectively.  Apart
from their negative norm, this added Higgs-doublet is a structure
which can be found in many extensions of the SM, most prominently
in the MSSM.  It is worth stressing at this point that the LW Higgs 
doublet $\tilde H$ does not aquire a vacuum expectation value.

\subsubsection{Neutral, CP-even Higgs bosons}
\label{sec:NeutralHiggsBosons}
The kinetic term for the neutral, CP-even Higgs bosons reads
\begin{equation}
{\cal L}_{h} = \frac{1}{2}\,
\left(\hat D_\mu\,{\cal H}\right)^\dagger\eta_2\,
\left(\hat D^\mu\,{\cal H}\right) - 
\frac{1}{2}\,{\cal H}^\dagger {\cal M}_h\eta_2\,{\cal H}\,,
\label{eq:higgskin}
\end{equation}
where the covariant derivative $\hat D_\mu$ is defined as
\begin{equation}
\hat D_\mu = 
\partial_\mu +i ({\bf A}_\mu + {\bf \tilde  A}_\mu) \,,
\label{eq:covariant}
\end{equation}
with the abbreviation ${\bf A_\mu} = g A_\mu^a T^a + g_2 W_\mu^a T^a
+ g_1 B_\mu\, Y$ for all SM gauge fields and $\tilde {\bf A}_\mu$ 
the analogous expression for the LW versions of the gauge fields.  
Furthermore, the two CP-even scalars of the theory, $h_0$ and
$\tilde h_0$ are now arranged in a vector ${\cal H}$, and they mix
\begin{equation} 
{\cal H} = \left( \bac h_0 \\ \tilde{h}_0 \ea \right)\,,
  \qquad {\cal M}_h\eta_2 = \frac{1}{2}\, \left(\bacc
    \lambda v^2 & -\lambda v^2 \\
    -\lambda v^2 & \lambda v^2-2 M^2_{H}\ea \right) 
\quad {\rm and} \quad 
\eta_2 \ =\ \left( \bacc 1 & 0\\0 & -1\ea\right)\,.
\end{equation}
The matrix ${\cal M}_h\,\eta_2$ can only be diagonalized if 
$M_H > \sqrt{2 \lambda}\,v$.  The diagonalization is achieved by the 
symplectic transformation $S_h$ satisfying
\begin{equation}
\label{eq:Hsym}
S_h \ =\ \left( \bacc
\cosh \phi_h & \sinh \phi_h \\
\sinh \phi_h & \cosh \phi_h \ea \right)\,,\qquad 
S_h \eta_2 S_h^\dagger = \eta_2 \,,
\end{equation}
where the fields ${\cal H}$ and the matrix ${\cal M}_h$ transform 
as
\begin{equation}
\label{eq:HT}
\physs{\cal H}  =  
\eta_2 S_h^\dagger \eta_2{\cal H}\,,  \qquad 
\phys{{\cal M}}{h}\,\eta_2 = 
S_h^\dagger\,{\cal M}_h\eta_2\,S_h \,.
\end{equation}
For the sake of notational brevity the suffix ``phys'' will be 
dropped later on for the fields but it will be retained on the 
matrices and the mass eigenvalues for clarity.  The symplectic 
rotation angle $\phi_h$ and the diagonalized mass matrix 
$\phys{\cal M}{h}$ are given by
\begin{equation}
\tanh 2\,\phi_h =
\frac{-\lambda\,v^2/M^2_H}{1-\lambda\,v^2/M^2_H}\,,
\end{equation}
and
\begin{equation}
\phys{\cal M}{h} \ = \left( \bacc
\frac12\left(M_H^2-\sqrt{M_H^4-2v^2\lambda  M_H^2}\right)&0\\
0 & \frac12\left(M_H^2+\sqrt{M_H^4-2v^2\lambda  M_H^2}\right)\\
\ea \right)\,,
\end{equation}
where the physical masses are the same as in the toy model 
Eq.~\eqref{eq:mMphys} with appropriate substitutions.
In terms of the physical neutral Higgs boson masses $\mph{h_0}$ and
$\mph{\tilde{h}_0}$ the quartic coupling $\lambda$ can simply be written
\begin{equation}
\label{eq:rhoHiggs}
\lambda\,v^2 \ =\ \frac{2\,\mph{h_0}^2\mph{\tilde h_0}^2}{\mph{h_0}^2 + \mph{\tilde
   h_0}^2}\,.
\end{equation}

Prior to diagonalization of the neutral Higgs boson mass matrix, from
Eq.~\eqref{eq:higgskin} notice that the LW Higgs fields do not couple to gauge
bosons via a trilinear coupling after electroweak symmetry breaking,
suggesting a ``gaugeophobic'' structure. This follows from the fact that the
$\tilde{H}$ does not aquire a VEV. However, after diagonalization, mixing will
induce these trilinear couplings with a strength proportional to the hierarchy
between the SM-like and LW Higgs masses.

\subsection{Fermions}

\subsubsection{Kinetic terms}
For the fermions, transforming between the higher derivative formalism and the
LW picture is slightly more involved. The higher derivative term Eq.\
\eqref{eq:HDfermion} generates a Dirac-type mass for the associated LW fields.
Therefore each SM fermion, treated as a massless chiral field before
electroweak symmetry breaking, receives one left-- and one right--handed LW
fermion partner. This results in an apparent tripling of the number of degrees
of freedom, rather than a doubling as seen in the scalar sector.
Alternatively, since they are charged, it is clear that the LW fermions need
to have a mass-term of the Dirac-type necessitating left-- and right--handed
fields of the same SU(2)$_L$ representation.

\subsubsection{Yukawa interactions and physical masses}
\label{sec:Yukawa}

After introducing the required auxiliary fields and performing a 
shift in the field variable, as described above, the Yukawa 
interactions for the quarks take the following form
\begin{equation}
{\cal L}_{\rm Yuk} \ =\ 
g_u^{ij}\,\big( \overline{u}_R^i - \overline{\tilde{u}}_R^i\big)\,
	  \big(H - \tilde{H} \big)\,\epsilon\,
          \big(Q_L^j - \tilde{Q}_L^j \big) \ - \ 
g_d^{ij}\,\big( \overline{d}_R^i - \overline{\tilde{d}}_R^i\big)\,
          \big(H^\dagger - \tilde{H}^\dagger \big)\,
          \big(Q_L^j - \tilde{Q}_L^j\big) \ +\ {\rm h.c.}\,.
\label{eq:yukint}
\end{equation}
In the lepton sector the Yukawa interactions also take a similar form. All
quarks apart from the top quark will be assumed to be massless, which is
expected to be a good approximation in the limit that the LW mass scales are
larger than the top mass $\mph{t}$. Only one Yukawa coupling will therefore be
considered, $g_u^{33} \simeq 1$. It will also be assumed that the LW mass
matrices of the $\tilde{Q}_L$ and $\tilde{u}_L$ are diagonal. Such an
assumption is compatible with the principle of Minimal Flavour Violation (MFV)
\cite{MFV}.
  
An additional Yukawa term coupling the right--handed SU(2)$_L$ doublet fermion
and the left--handed SU(2)$_L$ fermion singlet is allowed by the SU(2)$_L$
symmetry. This could have been written
\begin{equation}
\label{eq:Yukawa_ex}
\delta {\cal L}_{\rm Yuk} \sim 
\overline{ \tilde u_L'} (Y_H H + Y_{\tilde H} \tilde H) \epsilon Q_R' \,,
\end{equation}
but it is not required by the model. As pointed out in appendix
\ref{app:invariant}, this term does not have any impact on the results of our
paper and it is therefore neglected in the following.

For the diagonalization of the fermion mass matrix it is convenient 
to put each flavour into a three dimensional vector, such that 
\begin{equation}
\Psi^{t\,\top}_L = (
T_L,\tilde{T}_L, \tilde{t}^\prime_L)\,,\qquad \Psi^{t\,\top}_R = (
t_R,\tilde{t}_R, \tilde{T}^\prime_R) \,,
\end{equation}
where $T_L$ is a component of the third generation SM doublet $Q_L$
\begin{equation}
Q_{L\,3} = \left( \begin{array}{c} T_L \\ B_L \end{array} \right) \,,
\end{equation}
Lower case fermions denote SU(2)$_L$ singlets.  Note that each chiral fermion,
taking $T_L$ as an example, necessitates two chiral fermion partners, $\tilde
T_L$ and $\tilde T_R'$, to form the massive LW degree of freedom.  In this
formalism, the neutral Higgs-top interactions are given by
\begin{equation}
{\cal L} \ = \  
\frac{\sqrt{2}}{v} (h_0-\tilde h_0) \, \overline
{\Psi^t_R}\,g_t\,\Psi^t_L +\ {\rm h.c.}\,.
\label{topyukawa}
\end{equation}
Here, the parametrisation of the Higgs fields introduced in
Eq.~\eqref{eq:unitarygauge} has been employed and the Yukawa couplings are
encoded in the matrix
\begin{equation}
\label{eq:gt}
g_t \ =\  \left( \begin{array}{ccc}
m_t & -m_t & 0 \\
-m_t & m_t & 0 \\
0 & 0 & 0 \end{array} \right)\,.
\end{equation}
The presence of two LW fermion partners means the diagonalization of the
fermion mass matrices is slightly more involved than the scalar case
\cite{GOW}. The kinetic term assumes the following form
\begin{equation}
\label{eq:Lkin}
{\cal L}_{\rm kin} \ =\ 
\overline{\Psi^t}\,i\,\eta_3\,\fslash{\hat D}\,\Psi^t\ -\ 
\overline{\Psi_R^t}\,{\cal M}_t \eta_3 \,\Psi_L^t \ -\ 
\overline{\Psi_L^t}\,\eta_3 {\cal M}^{\dagger}_t  \,\Psi_R^t\,,
\end{equation}
with
\begin{equation}
{\cal M}_t\eta_3 \ =\ \left( \begin{array}{ccc}
m_t & -m_t & 0 \\
-m_t & m_t & -M_u \\
0 & -M_Q & 0 \end{array} \right)
\;\;\;\mbox{\rm and}\;\;\;
\eta_3 \ =\ \left( \begin{array}{ccc}
1 & 0 & 0 \\
0 & -1 & 0 \\
0 & 0 & -1 \end{array} \right)\,.
\label{topmass}
\end{equation}
Notice that Eq. \eqref{topmass} incorporates the fermion to SM gauge boson,
and fermion to LW gauge boson interactions via the covariant derivative $\hat
D$, defined in Eq.  \eqref{eq:covariant}. These interactions are essential for
the calculations in section \ref{sec:gluephoton}.

The mass matrix ${\cal M}_t$ can be diagonalized by separate left 
and right transformations $S_L$ and $S_R$ satisfying
\begin{equation}
\label{eq:property3}
S_L \eta_3 S_L^\dagger = \eta_3 
\;\;\;\mbox{\rm and}\;\;\;
S_R \eta_3 S_R^\dagger = \eta_3  \,.
\end{equation}
The fields $\Psi_{L(R)}$ and the matrices ${\cal M}_t$ and $g_t$
transform as
\begin{equation}
\label{eq:lorentz3}
\phys{\Psi}{L(R)}  =  \eta_3  S_{L(R)}^\dagger \eta_3  \Psi_{L(R)}
\,,\;\;\;
\phys{\cal M}{t} \eta_3 \ =\ S_R^\dagger\,{\cal M}_t \eta_3 \,S_L 
\;\;\;\mbox{\rm and}\;\;\;
\phys{g}{t} \ =\ S_R^\dagger\,g_t\,S_L\,.
\end{equation}
The $S_{L(R)}$ can be treated as two symplectic rotations and one unitary
transformation.  Notice that the matrix $\phys{g}{t}$ is not diagonal in
general.  In the same way as in the Higgs sector the ``phys'' symbol will be
omitted for the fermions for notational brevity but it will be retained on the
matrices and the physical masses for clarity.

Performing the matrix diagonalization analytically leads to lengthy
expressions.  Therefore, explicit results for the physical masses 
$\phys{\cal M}{t}$, the transformation matrices $S_{L,R}$ and the 
Yukawa couplings $\phys{g}{t}$ will be given as a series expansion 
in $\varepsilon \equiv m_t/M$, where the limit $M_Q = M_u \equiv M$
is explicitly used:
\begin{equation}
\phys{\cal M}{t} \ =\ M\,\left(
\begin{array}{ccc}
 \varepsilon + \varepsilon ^3 & 0 & 0 \\
 0 & 
 +1-\frac12\varepsilon - \frac38\varepsilon^2-\frac12\varepsilon^3 &
 0 \\
 0 & 0 & 
 +1 +\frac12\varepsilon -\frac38\varepsilon^2+\frac12\varepsilon^3
\end{array}
\right) \ +\ {\cal O}(M \varepsilon^4)\,,
\end{equation}
and
\begin{eqnarray}
\label{eq:SLSR}
S_L & = &
\left(
\begin{array}{ccc}
 1+\frac12\varepsilon ^2 & 
 \frac{1}{\sqrt{2}}\varepsilon +\frac{5}{4\sqrt{2}} \varepsilon^2 & 
 \frac{i}{\sqrt{2}} \varepsilon -\frac{5i}{4\sqrt{2}}\varepsilon^2 \\
 -\varepsilon ^2 & 
 -\frac{1}{\sqrt{2}}+\frac{1}{4\sqrt{2}}\varepsilon+
  \frac{1}{32\sqrt{2}}\varepsilon^2 &
  \frac{i}{\sqrt{2}}+\frac{i}{4\sqrt{2}}\varepsilon-
  \frac{i}{32 \sqrt{2}}\varepsilon^2 \\
 -\varepsilon & 
 -\frac{1}{\sqrt{2}}-\frac{1}{4\sqrt{2}}\varepsilon-
  \frac{15}{32\sqrt{2}} \varepsilon^2 &
 -\frac{i}{\sqrt{2}}+\frac{i}{4 \sqrt{2}} \varepsilon-
  \frac{15 i}{32\sqrt{2}}\varepsilon^2
\end{array}
\right)
\ +\ {\cal O}\left(\varepsilon^3 \right)\nonumber\\[5pt]
S_R & = & S^*_L\,.
\end{eqnarray}
Finally, the Yukawa couplings are proportional to the matrix
\begin{eqnarray}
\phys{g}{t} \ &=&\ M\,
\left(
\begin{array}{ccc}
 \varepsilon +3 \varepsilon^3 & 
 \frac{1}{\sqrt{2}}\varepsilon +\frac{3}{4\sqrt{2}} \varepsilon^2+
   \frac{87}{32 \sqrt{2}} \varepsilon^3 &
 -\frac{i}{\sqrt{2}} \varepsilon +\frac{3i}{4\sqrt{2}}\varepsilon^2-
   \frac{87i}{32\sqrt{2}} \varepsilon^3 \\
 \frac{1}{\sqrt{2}}\varepsilon +\frac{3}{4\sqrt{2}}\varepsilon^2+
   \frac{87}{32 \sqrt{2}} \varepsilon^3 & 
 \frac12\varepsilon+\frac34 \varepsilon^2+\frac32 \varepsilon^3 & 
 -\frac{i}{2} \varepsilon -\frac{15 i}{16} \varepsilon^3 \\
 -\frac{i}{\sqrt{2}}\varepsilon +\frac{3i}{4\sqrt{2}}\varepsilon^2-
  \frac{87 i}{32 \sqrt{2}} \varepsilon^3 & -\frac{i}{2}\varepsilon 
   -\frac{15 i}{16} \varepsilon^3 & 
 -\frac{1}{2}\varepsilon +\frac{3}{4}\varepsilon^2-
  \frac{3}{2} \varepsilon^3
\end{array}
\right) \nonumber\\
\ &+&\ {\cal O}\left(M\,\varepsilon^4\right)
\end{eqnarray}
It must be stressed here that the diagonalization of the mass terms is
possible only if $M\,>\,3 \sqrt{3}\,m_t/2$. The cubic equation for the
physical masses of the top quarks is given in appendix \ref{app:matrices}.
Also given in appendix \ref{app:matrices} are expressions for the matrices
$\phys{g}{t}$ and $\phys{\cal M}{t}$ in an expansion more closely related to
physical masses.

The simple relation between $S_R$ and $S_L$ in Eq.~\eqref{eq:SLSR} is due to
the fact that for $M_Q = M_u$ the matrix ${\cal M}_t \eta_3$ is symmetric.
This in turn implies that the property $g_t^\top = g_t$ is preserved by the
transformation \eqref{eq:lorentz3} to the physical matrix $\phys{g}{t}$.

Note that the diagonalization of fermions does not affect the
coupling to the gauge bosons in Eq.~\eqref{eq:Lkin} due to
the defining property \eqref{eq:property3}.

\subsection{Gauge Bosons}
\label{sec:bosons}

\subsubsection{One gauge field}

The gauge sector is slightly more involved than the fermion and scalar
sectors. As a ``warm-up'' exercise, first a single gauge sector will be
discussed.  In this discussion it will become apparent that the auxiliary
degree of freedom introduced to deal with the higher derivatives is not a
gauge field -- the new fields have a mass-term from the start and hence they
are truly massive spin-1 fields, i.e.\ Proca fields.

The gauge sector in the higher derivative formulation is
defined through the usual gauge kinetic term 
$-1/2  {\rm tr}  \big(\hat F_{\mu\nu}\hat F^{\mu\nu} \big)$ plus the 
higher derivative kinetic term given in Eq.~\eqref{eq:HDvector}, 
which generates new types of interaction terms.  The higher derivative 
term can be eliminated by introducing the auxiliary field 
$\tilde A_\mu$ \cite{GOW}
\begin{equation}
{\cal L}_{gauge} = 
-\frac{1}{2} {\rm tr} \big(\hat F_{\mu\nu}\hat F^{\mu\nu} \big) 
-M_A^2 {\rm tr} \big(\tilde A_\mu \tilde A^\mu\big)
+ 2  {\rm tr} \big(\hat F_{\mu\nu}\hat D^\mu \tilde A^\nu  \big) \, ,
\end{equation}
where 
$\hat D_\mu \tilde A_\nu = 
\partial_\mu \tilde A_\nu - ig [\hat A_\mu,\tilde A_\nu]$. Clearly, for an
Abelian gauge symmetry the commutator of the gauge fields would vanish. As
before, the higher-derivative Lagrangian is obtained by eliminating the
auxiliary field by its EoM.  Moreover the EoM
\begin{equation}
 (\hat D^\mu \hat F_{\mu\nu})^A + M_A^2 (\tilde A_\nu)^A = 0  
\end{equation}
explicitly shows that the auxiliary is \emph{not} a gauge field.   
Under a local gauge transformation $G(x)$ the two fields 
$\hat A_\mu$ and $\tilde A_\mu$ transform as
\begin{eqnarray}
\label{eq:gT}
\hat  A_\mu  &\to&  
G(x) \left(\hat  A_\mu + \frac{i}{g} \partial_\mu \right) G(x)^{-1}
\nonumber \\\
\tilde  A_\mu  &\to&  G(x)  \tilde  A_\mu  G(x)^{-1} \,.
\end{eqnarray}
The kinetic terms are diagonalized by the field shift 
$\hat A_\mu  = A_\mu + \tilde A_\mu$, where $A_\mu$ transforms as 
$\hat A_\mu$ \eqref{eq:gT} under a local gauge transformation.   
The Lagrangian then assumes the form
\begin{eqnarray}
\label{eq:gauge}
{\cal L}_{gauge} =& & 
-\frac{1}{2} {\rm tr} \big( F_{\mu\nu}F^{\mu\nu} \big) +  
\frac{1}{2} {\rm tr}\Big(  
		(D_\mu \tilde A_\nu  \! - \! D_\nu \tilde A_\mu) 
		(D^\mu \tilde A^\nu  \! - \! D^\nu \tilde A^\mu) 
	\Big) 
\nonumber \\
& &  
-M_A^2  {\rm tr} \big(\tilde A_\mu \tilde A^\mu\big) +
{\rm tr} \left( 
[\tilde A_\mu, \tilde A_\mu] 
	\Big(-i g F^{\mu\nu} -\frac{3}{2} g^2 [\tilde A^\mu, \tilde A^\mu] 
	-4 i gD^\mu \tilde A^\nu \Big)  
\right) \,.
\end{eqnarray}
Note that there are no terms linear in either $A_\mu$ or 
$\tilde A_\mu$, as demanded by the diagonalization, which is not 
immediately obvious from the expression above.
The propagators for the fields are
\begin{eqnarray}
\label{eq:gaugeprop}
D_{\mu\nu}(k)  &=& 
\frac{i}{k^2}\Big( g_{\mu\nu} - (1-\xi) \frac{k_\mu k_\nu}{k^2} \Big) 
\nonumber  \\
D_{\mu\nu}(k)  &=& 
\frac{i}{k^2-M_A^2}\Big( g_{\mu\nu} - \frac{k_\mu k_\nu}{M_A^2} \Big) 
\,.
\end{eqnarray}
The standard $R_\xi$ gauge has been chosen for the gauge field $A_\mu$.  As
already emphasized, the field $\tilde A_\mu$ is not a gauge field, its
propagator is that of a massive vector field, formally identical to that of,
say, the $W$ gauge boson propagator in unitary gauge.  This fact will be
further exploited in section \ref{sec:gaga} when calculating the $W$ loop
contribution to $h_0 \to \gamma\gamma$.

\subsubsection{The complete SM gauge sector}
In this section, the mass diagonalization of the gauge sector in the LWSM will
be addressed. After electroweak symmetry breaking, the masses of the
electroweak gauge bosons have the following form
\begin{eqnarray}
{\cal L}_{\rm gauge} & = & 
\frac12{\cal B}^\top_\mu\,{\cal M}_{\cal B}\,\eta_4\,{\cal B}^\mu
\ +\ 
\frac12{\cal W}^{a\top}_\mu\,{\cal M}_{\cal W}\,\eta_2\,
        {\cal W}^{a\,\mu}\,,
\quad {\rm where}
\quad
\eta_4 \ =\ {\rm diag}(1,1,-1,-1)\,,\nonumber\\[5pt]
{\cal M}_{\cal B}\,\eta_4 & = & 
\left( \begin{array}{lc}
M_{\rm SM}\hspace*{5mm} & M_{\rm SM}\\
M_{\rm SM}\hspace*{5mm} & M_{\rm SM} - M_{12} \end{array} \right)\,,
\quad 
M_{\rm SM} \ =\ \frac{v^2}{4} \left( \begin{array}{cc}
g_1^2 & -g_1\,g_2\\
-g_1\,g_2 & g_2^2 \end{array} \right)\,,\nonumber\\[5pt]
{\cal M}_{\cal W}\,\eta_2 & = & 
\frac{1}{4} \left( \begin{array}{cc}
g_2^2\,v^2\hspace*{5mm} & g_2^2\,v^2\\
g_2^2\,v^2\hspace*{5mm} & g_2^2\,v^2 - 4 M_2^2\end{array} \right)\,,\quad
M_{12} \ =\ \left( \begin{array}{cc}
M_1^2 & 0\\
0 & M_2^2 \end{array} \right)\,.
\end{eqnarray}
In the equations above, $a = \{1,2\}$, 
${\cal B}_\mu^\top=(B_\mu,\,W^3_\mu,\,\tilde B_\mu,\,\tilde W^3_\mu)$,
${\cal W}_\mu^{a\,\top} = ( W^a_\mu, \tilde{W}^a_\mu )$ and $M_1$, 
$M_2$ are the LW masses in the gauge boson sector.

The 2$\times$2 matrix ${\cal M}_{\cal W}\, \eta_2 $ is diagonalized 
by a symplectic transformation such that 
\begin{equation}
\label{eq:WT}
\physs{\cal W} \ =\ \eta_2\,S_W^\dagger\,\eta_2\,{\cal W}\,.
\end{equation}
The matrix $S_W$ satisfies
\begin{equation}
S_W \ =\ \left( \bacc
\cosh \psi_W & \sinh \psi_W \\
\sinh \psi_W & \cosh \psi_W \ea \right)\,,\qquad 
S_W \eta_2 S_W^\dagger = \eta_2 \,,
\end{equation}
where
\begin{equation}
\tanh 2\,\psi_W \ =\ \frac{g_2^2\,v^2}{2M_2^2-g^2_2\,v^2}\,,
\end{equation}
and
\begin{eqnarray}
\label{eq:mW}
\mph{W}^2 & = & \frac12\left(M_2^2-\sqrt{M_2^4 -
    g_2^2\,v^2\,M_2^2}\right)\,,\nonumber\\
\mph{\tilde{W}}^2 & = & \frac12\left(M_2^2+\sqrt{M_2^4 -
    g_2^2\,v^2\,M_2^2}\right)\,.
\end{eqnarray}

The 4$\times$4 matrix ${\cal M}_{\cal B}\, \eta_4$ is diagonalized by 
a combination of symplectic and orthogonal transformations such that 
mixing between the upper two and lower two components occurs via a 
symplectic rotation and mixing amongst either the upper or lower two 
components occurs via an orthogonal rotation.  Under this 
transformation,
\begin{equation}
\physs{\cal B} \ =\ \eta_4\,S_B^\dagger\,\eta_4\,{\cal B}\,, 
\end{equation}
and one physical mass is guaranteed to be zero (corresponding to 
the photon).

In the following, it will be assumed that mixing between the SM and LW sectors
can be treated as a perturbation to the usual SM mixing \cite{GOW}.  This
allows the upper 2$\times$2 block of ${\cal M}_{\cal B}$ to be diagonalized by
an orthogonal rotation about an angle $\theta_W$, the usual weak mixing angle.
The LW fields can then be diagonalized by another orthogonal rotation with a
mixing angle
\begin{equation}
\tan 2 \phi_W = \frac{g_1\,g_2\,v^2}{2}\,\left(M_1^2 - M_2^2 + (g_1^2 -
  g_2^2)\,\frac{v^2}{4} \right)^{-1}\,.
\end{equation}
Hence, depending on the actual values of $M_1$ and $M_2$ this angle
can assume any value and for $M_1 = M_2$ it is, not surprisingly,
identical with the original Weinberg angle. 
 
As pointed out by GOW \cite{GOW}, mixing between the neutral SM and LW gauge
bosons will lead to a tree-level contribution to the electroweak $\rho$
parameter. Bounds on $\Delta \rho$ place constraints on $M_1$ such that $M_1
\stackrel{>}{{}_\sim} 1$~TeV, thus placing bounds on the masses of the neutral
LW gauge bosons. Clearly a more careful analysis would be desirable, but for
the purposes of this work $M_1 >1$~TeV will be assumed.  It should be pointed
out that measurements of muon decay do not set a tree level constraint on the
the mass of the LW $W$ boson, contrary to models with a $W'$ \cite{PDG}. This
can for instance be seen in the higher derivative formulation by observing
that the propagator \eqref{eq:prop} at zero momentum transfer $\hat D(0) =
-i/m^2$ does not contain any trace of the LW mass scale!  Bounds on the LW $W$
boson are therefore expected to be less stringent than those on the LW $Z$
boson. Bounds from direct searches for $W'$s and $Z'$s are model dependent,
but are typically in the range 780-920~GeV for heavy SM-like gauge bosons
\cite{directsearch}.

\section{Higher derivatives vs LW formulation}
\label{sec:quads}

In this section the mechanism responsible for cancelling quadratic divergences is outlined in both the higher derivative formalism, as advocated in Ref.~\cite{GOW}, and the original LW formulation 
\cite{finiteqed}. Moreover,
at the end of this section some comments on the consequences for finite graphs
are added.

Introducing higher derivatives improves the convergence of graphs by adding
higher powers of momenta in the propagators. This works as long as potential
higher derivatives in the interaction terms do not upset this effect. In the
LW model higher derivatives are introduced into the kinetic terms but also to
the interaction terms, e.g.\ $\hat \phi^\dagger \hat D^4 \hat \phi$
\eqref{eq:HDscalar}, through the principle of minimal coupling. It is
therefore not obvious that the Higgs boson self-energy contribution due to
gauge fields has any improved convergence. It was pointed out in \cite{GOW},
by assuming the Landau gauge, that the potentially dangerous derivatives in
the interaction term can be moved to the external legs.  This assures improved
convergence, which means that the quadratic divergences are absent.  The
absence of quadratic divergences in the Higgs boson self energy due to the
fermions is immediate in the higher derivative formulation, contrary to the
case discussed above, since the Yukawa interaction term is not affected by
additional derivatives enforced by a minimal coupling.

It is also instructive to investigate the cancellation of quadratic
divergences in the Higgs boson self energy due to fermions in the LW
formulation.  Naively, it is not obvious how this happens, but since the two
formulations are equivalent for physical quantities it is clear that
cancellations must take place.

After diagonalization of the fermion states there are nine possible graphs per
flavour corresponding to all different pairings of propagators of the fermion
states $\Psi^{t\,\top}_{L(R)} = (T_{L(R)},\tilde{T}_{L(R)},
\tilde{t}^\prime_{L(R)})$.  The quadratically divergent part is proportional
to
\begin{equation}
\Sigma_H|_{\rm div} \sim \Lambda_{\rm cut-off}^2 {\rm Tr}[\eta_3\,
\phys{g}{t}\, \eta_3 \,\phys{g}{t}^\dagger] = \Lambda_{\rm cut-off}^2
 \sum_{i,j}  |(\phys{g}{t})_{ij}|^2 (-1)^{\delta_{i1}} (-1)^{\delta_{j1}}
 \,,
\end{equation}
which is invariant under the orthosymplectic  transformation
Eq.~\eqref{eq:lorentz3}
\begin{equation}
I_{g_t} =  {\rm Tr}[\eta_3\, \phys{g}{t}\, \eta_3 \,\phys{g}{t}^\dagger] =
 {\rm Tr}[\eta_3\,  g_t\, \eta_3 \, g_t^\dagger]  \, .
\end{equation}
This is easily verified by the use of the property Eq.~\eqref{eq:property3}.
Note that the invariant properly takes into account the minus signs of the
ghosts.  Therefore Eq.~\eqref{eq:gt} implies that the invariant $I_{g_t} =
|m_t|^2 ( 1+1 -1-1) = 0$ vanishes guaranteeing the absence of quadratic
divergences.  A quicker way to see this, which has already been advocated in
reference \cite{GOW}, is to use the fact that the quadratic divergencies are
independent of the mass and that therefore the masses can be set to
zero\footnote{ The limit to zero mass is continuous for fermions and scalars
  used here, whereas for vectors this is not the case and additional
  subtleties could arise.}  in which case the states are diagonal from the
beginning and the absence of quadratic divergencies again follows from
Eq.~\eqref{eq:gt} and the ghosts signs.

It was shown above how quadratic divergences cancel, even in the slightly
complicated case of the fermion loop.  It is worth pointing out that the
quadratic divergences may re-enter via the back door through the mass of the
virtual LW particle in cases where the diagrams do not decouple in the
infinite mass limit, which is the case for the quadratically divergent
subprocesses.  A simple an example is the quartic Higgs boson self interaction
for which the corresponding self energy is
\begin{equation}
\Sigma_H \sim \frac{ \lambda}{4 \pi^2}  \Big\{  (\Lambda_{\rm cut-off}^2 +
c\, \mph{h_0}^2 + \dots )  -  (\Lambda_{\rm cut-off}^2 + c \,\mph{\tilde
h_0}^2 + \dots )  \Big\} \,,
\end{equation}
where the dots stand for logarithmic corrections.  The meaning of this
intuitive result is that in order to solve the hierarchy problem the LW Higgs
mass should not be much higher than the electroweak scale. 

Furthermore, as stated in the introduction it was shown in 
Ref.~\cite{nuR_hierarchy} that in the context of the see-saw mechanism a heavy neutrino can be
embedded into the model without destabilizing the Higgs mass.
More precisely a correction of the form
$\delta m_{h_0}^2 \sim m_{\nu_R}^2$, which is caused in the 
see-saw extended SM through the loop of the left handed and right handed neutrino, does not appear in the see-saw extension of the 
LWSM, as long as $m_{\nu_R} \gg m_{\nu_L}$.

 After
understanding what happens to processes which are divergent in the SM, it is
natural to ask what the impact is on processes which are \emph{finite} in the
SM.  From the above discussion and the explicit form of the propagators
\eqref{eq:prop} it could be anticipated that the higher derivative formulation
implies corrections of the form $ (\sqrt{s}/M_{LW})^n$ with $s$ the energy of
the process and $n$ some positive integer power.  In the LW formulation the
same result follows from the viewpoint of the Appelquist--Carazzone decoupling
theorem \cite{decoupling}.

\section{The process $gg \to h_0 \to \gamma\gamma$ in the Lee-Wick SM}
\label{sec:gluephoton}

The discussion of the process $gg \to h_0 \to \gamma\gamma$ is one of the main
topics of this paper since it has a huge importance for Higgs boson searches,
especially for comparably low Higgs boson masses.  The main Feynman graphs at
leading order are depicted in Fig.~\ref{fig:gluonphoton}.  It is useful to
parametrize the ratio of amplitudes in the LWSM to the amplitudes in the SM as
\begin{equation}
\label{eq:kappa}
\kappa_{gg} = \frac{{\cal A}_{\rm LW}(gg\to h_0)}{{\cal A}_{\rm SM}(gg\to
  h_0)}\,,\quad {\rm and}\quad
\kappa_{\gamma \gamma} = \frac{{\cal A}_{\rm LW}(h_0\to \gamma\gamma)}{{\cal
    A}_{\rm SM}(h_0\to \gamma \gamma)}\,.
\end{equation}
The ratio of the cross-section times decay rate for the full process $gg\to
h_0 \to \gamma \gamma$ therefore reads
\begin{equation}
\frac{\sigma_{\rm LW}(gg \to h_0)\,\Gamma_{\rm LW}(h_0 \to \gamma \gamma)}
  {\sigma_{\rm SM}(gg \to h_0)\,\Gamma_{\rm SM}(h_0 \to \gamma \gamma)} \ =\ |\kappa_{gg}|^2\,|\kappa_{\gamma\gamma}|^2\,.
  \label{eq:sigmagamma}
\end{equation}
Differences between the SM and the LWSM show up in the process $gg \to h_0 \to
\gamma\gamma$ because of additional degrees of freedom propagating in loops
and also because of effects due to the mixing of the SM and LW fields e.g.
Eq.  \eqref{eq:HT},
\begin{eqnarray}
h_0  &=& \cosh \phi_h \, \phys{h}{0} + \sinh  \phi_h\, \phys{\tilde h}{0} \,
\nonumber \\
\tilde h_0  &=& \sinh \phi_h \, \phys{h}{0} + \cosh  \phi_h\, \phys{\tilde h}{0}\,.
\end{eqnarray}
The  mixing factors, which will be denoted by the letter $s$, are easily
read off from  Eq.~\eqref{eq:Hsym} and \eqref{eq:HT}\footnote{
	Note that the symplectic mixing angle $\phi_h$ is negative.}:
\begin{eqnarray}
\label{eq:s}
s_H &=& \cosh \phi_h  =  \frac{1}{( 1 - r_{h_0}^4  )^{1/2} }\,, 
\qquad \quad r_{h_0} \equiv \frac{\mph{ h_0}} {\mph{\tilde h_0}}\,,
 \nonumber  \\
s_{H\!-\!\tilde H} &=& \cosh \phi_h  - \sinh \phi_h =
\frac{ 1 + r_{h_0}^2  }{( 1 - r_{h_0}^4  )^{1/2} } 
 \, ,
\end{eqnarray}
where the subscripts denote the type of interaction coupling. Yukawa
interactions couple the fermions to the combination $H - \tilde H$ (c.f. Eq.
\eqref{eq:yukint}) and so the correction to this vertex is denoted
$s_{H-\tilde H}$. As discussed in section \ref{sec:NeutralHiggsBosons}, the LW
Higgs does not couple to the gauge bosons via a trilinear coupling in the
interaction basis, and therefore the Higgs-to-gauge boson vertex is scaled by
a factor $s_{H}$.

\subsection{$gg \to h_0$}
\label{sec:gg}

\begin{figure}[t]
\begin{center}
\includegraphics[width=10cm]{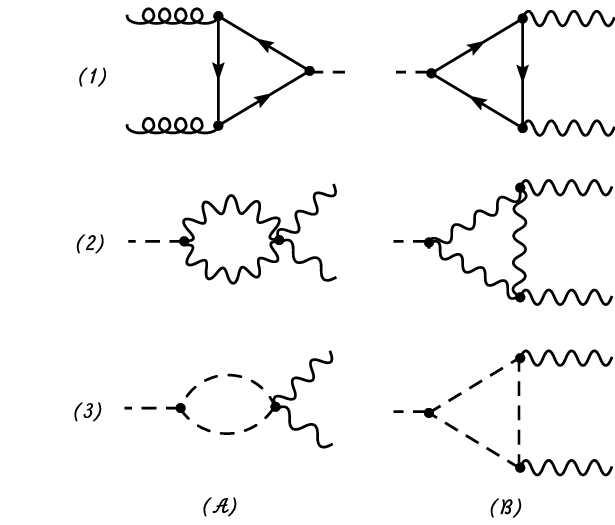}
\caption{\small\em 
	(1A) fermion loop to $gg \to h_0$. (1B) fermion, 
	(2A,2B) $W$ boson and (3A,3B) would-be Goldstone boson (SM) 
	or charged Higgs boson (LW) contribution to $h_0 \to \gamma \gamma$. 
	In the SM, the dominant contribution to $gg \to h_0$ stems from the 
	top quark and the dominant contribution to $h_0 \to \gamma\gamma$ is 
	due to the $W^\pm$.}
\label{fig:gluonphoton}
\end{center}
\end{figure}

The process $gg \to h_0$ is mediated by a fermion triangle loop as depicted in
Fig.~\ref{fig:gluonphoton}(1A).  In the SM the top contribution is the only
relevant contribution, c.f.  appendix \ref{app:triangle}.  The graph
essentially counts the number of heavy quarks $\mph{Q} \simge \mph{t}$.  The
fermion loop without couplings scales as $1/m_f$ in the heavy mass limit,
obeying the decoupling theorem. The total vertex is independent of the fermion
mass since the Yukawa coupling is proportional to $m_f$.  From the discussion
in the previous section the crucial question is what the power correction
$m_{h_0}/\mph{t,\tilde t,\tilde t'}$ is. The analysis will be first outlined
in the LW formulation, and then the argument will be given in the higher
derivative formulation.

After noticing how the triangle graph scales with the fermion mass, it is
clear that the the effective vertex in the LW formulation is proportional to
\begin{equation}
\Gamma_{ggh_0} \sim   {\rm Tr}[ \, \phys{g}{t} \, \eta_3\,  \phys{{\cal
M}}{t}^{-1}\,]
\,\, + \,\, O(1/\beta_t) ,
\end{equation}
where $\beta_t = 4 \mph{t,\tilde t,\tilde t'}^2/\mph{h_0}^2$ is the threshold
parameter. Note that the first term is independent of the actual values of the
various top quark masses and the remainder stands for the asymptotic
corrections to the $m_t \to \infty$ limit.  The leading term is invariant
under the orthosymplectic transformation of Eq.~\eqref{eq:lorentz3}.  Moreover
it has the additional curious property,
\begin{equation}
\label{eq:It}
I_t =  {\rm Tr}[ \, \phys{g}{t} \, \eta_3 \, \phys{{\cal M}}{t}^{-1}\,]  
=   {\rm
Tr}[ \,  g_t \, \eta_3 \,  {\cal M}_t^{-1}\,]  = 1  \,,
\end{equation}
of being equal to one, which is explicitly demonstrated in appendix
\ref{app:invariant}. Furthermore, the invariant remains unity even if a Yukawa
term of the form Eq.~\eqref{eq:Yukawa_ex} would be added; again this finding
is discussed in more detail in appendix \ref{app:invariant}.

Another way to understand this result is by inspecting the fermion propagator
of the higher derivative formulation.  The additional term in
Eq.~\eqref{eq:HDfermion} leads to a propagator
\begin{equation}
\hat S_F(p) = \frac{\fslash{p}\big(1-\frac{p^2}{M^2}\big) +
m}{p^2(1-\frac{p^2}{M^2}\big)^2 -m^2} \, .
\end{equation}
Notice that the graph remains finite because the increase in convergence
coming from the fermion propagators is sufficient to compensate for
higher-derivative interaction terms of the form $\delta {\cal L}_{hd} \sim
\bar t (\partial^2/M^2) \fslash{A} t$. The fermion triangle graph has a $SVV$
structure and the trace over the Dirac matrices vanishes unless the SM fermion
mass term $m$ is picked up. This means by power counting that the correction
from the LW fermions is of the order of ${\cal O}(s/M^2)$ and therefore the
power is $n = 2$.  The scaling arguments in the SM would in principle allow
for $n=1$. In the higher-derivative formulation this does not happen because
the LW fermion mass $M$ does not enter with a term proportional to the unit
matrix, in contrast to the SM fermion mass $m$. Notice that the difference in
the structure of the SM and LW mass terms is transparent in the higher
derivative formalism but not in the LW formalism. In the LW formulation the
scaling of the pure loops is $1/m_f$, as stated above, but the diagonalization
of the Yukawa matrix exactly cancels this effect. Furthermore, this implies
that the Yukawa element $(\phys{g}{t})_{11} > \mph{t}$ which is unusual and
due to the ghost-like nature of the LW particles.  The consequences for the
CKM elements $|V_{\rm t(d,s,b)}|$ will be discussed later, in section
\ref{sec:flavour}.

Finally, the $\kappa$ factor for the process $gg \to h_0$ is then simply given
by \footnote{The quantity $\tilde F_{1/2}$ may be written in terms of a trace
  as: $\tilde F_{1/2} = {\rm Tr}[ \, \phys{g}{t} \, \eta_3 \, \phys{{\cal
      M}}{t}^{-1} \, F \,]$ with $F = {\rm diag}[F_{1/2}(\beta_t) ,
  F_{1/2}(\beta_{\tilde t}) , F_{1/2}(\beta_{\tilde t'})]$.}
\begin{eqnarray}
\label{eq:kappagg}
\kappa_{gg} =
\frac{ s_{H\!-\!\tilde H} \, \tilde  F_{1/2} }{   F_{1/2}(\beta_{t})  }   
\,,
\end{eqnarray}
with
\begin{equation}
\label{eq:Ftilde}
\tilde  F_{1/2} =
 \frac{( \phys{g}{t})_{11}}{\mph{t}}     \,
F_{1/2}(\beta_{t}) -
  \frac{( \phys{g}{t})_{22}}{ {\mph{\tilde t}}} \, F_{1/2}(\beta_{\tilde
t}) -
  \frac{( \phys{g}{t})_{33}}{ {\mph{\tilde t'}}}   \,
F_{1/2}(\beta_{\tilde t'})
 \,,
\end{equation}
where $s_{H - \tilde H}$ takes into account the mixing of the SM and LW Higgs and is given 
in Eq.~\eqref{eq:s}. The the threshold parameter $\beta_x$ is
\begin{equation}
\beta_x = \frac{4 \mph{x}^2}{\mph{h_0}^2}\, .
\end{equation}
The form factor $F_{1/2}$ is given in the appendix \ref{app:triangle}
Eq.~\eqref{eq:F}.

\subsection{$h_0 \to \gamma\gamma$}
\label{sec:gaga}

In addition to the top quark loop, already discussed for the process $gg \to
h_0$, the process $h_0 \to \gamma\gamma$ receives further contributions from
the $W$ boson loop Fig.~\ref{fig:gluonphoton}(2A,2B) and LW charged Higgs
bosons Fig.~\ref{fig:gluonphoton}(3A,3B).

The dominant contribution to the process $h_0 \to \gamma \gamma$ comes
from charged gauge boson loops, as can be seen in appendix
\ref{app:triangle}. The gauge boson sector of the LW model is studied in 
section \ref{sec:bosons}.  A central finding there was that the LW gauge 
boson propagator \eqref{eq:gaugeprop} has a form identical (up to a sign) 
to the SM gauge boson propagator in the unitary gauge. Furthermore, all
relevant vertices are also identical up to signs.  Therefore, in the unitary 
gauge the LW $\tilde W$ boson contribution has an identical form to the SM 
$W$ boson up to some multiplicative factors.  Since the amplitude is 
independent of the choice of gauge this implies that the LW $W$ boson 
contribution is finite, although in naive power counting it is not finite.
This has been checked through an explicit calculation.  Moreover, further 
graphs containing mixed vertices of the type $\partial \tilde h^+ h^- A$ 
can be avoided by choosing the unitary gauge for the SU(2)$_L$ gauge
field which decouples the would-be Goldstone bosons from the Lee-Wick
charged Higgs bosons.

Taking everything together, the total correction factor $\kappa_{\gamma\gamma}$ 
is thus given by
\begin{equation}
\label{eq:kappaph}
\kappa_{\gamma\gamma} =
\frac{s_{H\! -\! \tilde H} \,(N_c Q_t^2) \tilde F_{1/2} +
      s_{H} \, \tilde F_1 +
      s_{H \!-\! \tilde H}\,
		\,\tilde F^\varphi_{0}}
     {(N_c Q_t^2) F_{1/2}+ F_{1} }  \,,
\end{equation}
where the form factors $F_{1/2}$ and $F_1$  are given in
appendix \ref{app:triangle} and the tilde form factors shall be discussed below.  The SM and LW Higgs mixing factors $s_{H -\tilde
  H}$ and $s_{H}$ are given in Eq.~\eqref{eq:s}.  In contrast to the SM, the
LW model has an additional contribution from charged LW Higgs bosons running
in the loop\footnote{ The form of their loop factor, of course, is identical
  to that of the would-be Goldstone bosons of the SM in non-linear $R_\xi$
  gauge.}.  Their Feynman rules may be read off from the Lagrangian in
Eq.~\eqref{eq:higgskin} and the parametrisation \eqref{eq:unitarygauge}.  The
top fermion contributions $\tilde F_{1/2}$ are the same as in the previous
section Eq.~\eqref{eq:Ftilde}.
As discussed above, the LW $\tilde W$ boson contribution
is the same as in the SM up to factors which are easily read off from the
Lagrangian.  There is also an effect due to the mixing of the $W$ bosons
Eq.~\eqref{eq:WT}
\begin{eqnarray}
\label{eq:WTex}
W  &=& \cosh \psi_W \, \physs{W} + \sinh  \psi_W\, \physs{\tilde W} \,,
\nonumber \\
\tilde W  &=& \sinh \psi_W \, \physs{W} + \cosh  \psi_W\, \physs{\tilde W} \,,
\end{eqnarray}
which results in a correction factor $s_{(A+\tilde A)^2}$, similar to
$s_H$ and $s_{H-\tilde H}$ in Eq.~\eqref{eq:s}, for the $h_0WW$
and $h_0\tilde W \tilde W$ vertices
\begin{equation}
\label{eq:sA}
s_{(A+\tilde A)^2} =( \cosh \psi_W + \sinh \psi_W)^2 = 
\frac{1 + r_W^2}{1-r_W^2}\,, \qquad \quad r_W \equiv 
\frac{\mph{W}}{\mph{\tilde W}} \, .
\end{equation}
The full $W$ boson contribution is then given by
\begin{equation}
\tilde F_{1} =
 s_{(A + \tilde A)^2}  \rho_{vW}  \left[ F_1(\beta_W)  -   
\left(\frac{ \mph{W}}{\mph{\tilde W}} \right)^2  F_{1}(\beta_{\tilde W}) \right]  \, ,
\end{equation}
where the $\rho_{vW}$ factor will be explained at the end of this section.
The correction $s_{(A + \tilde A)^2}$ to $\tilde F_1$ is analogous to the
correction $(\phys{g}{t})_{11}/\mph{t} \dots$ to $\tilde F_{1/2}$ in  Eq.
\eqref{eq:Ftilde}.
The combined pairs of vertices $(WWA,\tilde W \tilde WA)$ and
$(WWAA,\tilde W \tilde WAA)$, appearing in the diagrams shown in
Fig.~\ref{fig:gluonphoton}(2A,2B), are left unchanged under the mixing. From a
formal point of view this has to be the case since the triangle graph with the
first two couplings (Fig.~\ref{fig:gluonphoton}(2B)) and the self-energy
graphs (Fig.~\ref{fig:gluonphoton}(2A)) are not separately gauge invariant.
This effect is achieved by the fact that the vertex pairs have opposite signs
and substitution of Eq.~\eqref{eq:WTex} gives an overall factor of,
$\cosh\psi_W^2-\sinh\psi_W^2 = 1$, unity.
Finally the scalar form factor is given by
\begin{equation*}
\tilde F_{0}^\varphi  =  - \rho_{vH} 
\Big( \frac{\mph{h_0}}{\mph{\tilde h+}} \Big)^2  \frac{F_{0}^\eta(\beta_{ \tilde h+})}{2}\,.
\end{equation*}
The factors of $\rho_{vH,vW}$ stand for the mass ratios
\begin{alignat}{2}
\rho_{vH}  &=  \frac{(\lambda^2 v^2/2)}{ \mph{h_0}^2}  = \frac{1}{1+r_{h_0}^2}  \;, \quad  & r^2_{h_0} = \frac{\mph{h_0}^2}{\mph{\tilde  h_0}^2}  \;,  \nonumber \\[0.1cm]   \qquad \rho_{vW} &=   \frac{(g^2 v^2/4)}{ \mph{W}^2}  = \frac{1}{1+r_W^2}  \;, \quad  & r^2_W = \frac{\mph{W}^2}{\mph{\tilde W}^2}\;,
\end{alignat}
of Standard Model versus physical Lee-Wick masses of the Higgs and
the $W$ boson respectively and can be obtained from 
Eq.~\eqref{eq:rhoHiggs} and by inverting Eq.~\eqref{eq:mW}.  They take care of the fact that the 
amplitudes decouple as $1/m^2_{\rm loop}$ i.e. the inverse of the squared mass
of the loop particle.

\subsection{Quantitative analysis}
\label{sec:numerics}

\begin{figure}[t]
\begin{center}
\includegraphics[width=10.0cm]{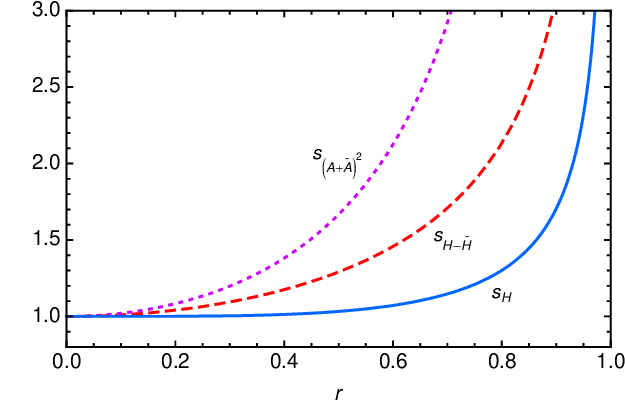}
\end{center}
\caption{\small \em Plots of the dimensionless correction factors,
  $s_{H}(r_{h_0})$, $s_{H-\tilde H}(r_{h_0})$ \eqref{eq:s} and $s_{(A +
    \tilde{A})^2}(r_{W})$ \eqref{eq:sA} as a function of $r_{h_0} \equiv
  \mph{h_0}/\mph{\tilde{h}_0}$ and $r_W \equiv \mph{W}/\mph{\tilde W}$.}
\label{fig:rscale}
\end{figure}

\begin{figure}[t]
\begin{center}
\begin{minipage}{11.5cm}
\begin{flushright}
\includegraphics[width=10.0cm]{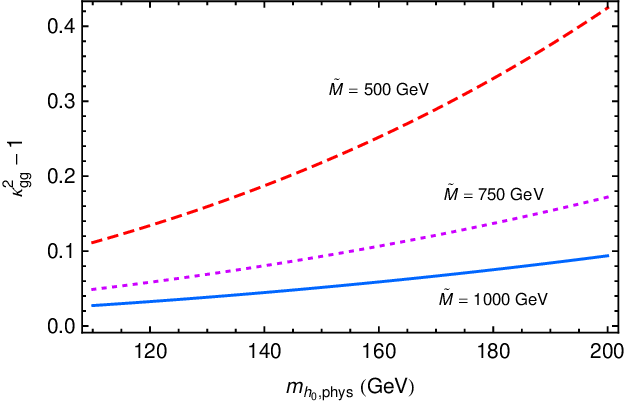}\\[0.6cm]
\includegraphics[width=10.7cm]{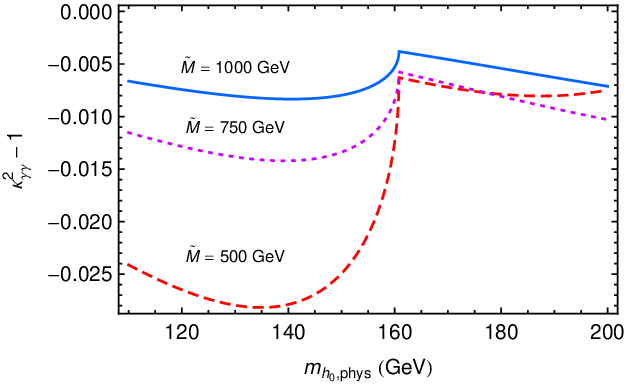}
\end{flushright}
\end{minipage}
\end{center}
\caption{\small \em The relative changes in the rates for $gg \to h_0$ and
  $h_0 \to \gamma \gamma$ in the LWSM, expressed as $|\kappa_{gg}|^2 -1$ and
  $|\kappa_{\gamma\gamma}|^2 -1$ respectively, plotted as a function of
  $\mph{h_0}$. Lee-Wick mass scales are such that $M_Q = M_u = \mph{\tilde{h}}
  = \mph{\tilde{h}_+} = \mph{\tilde{W}} \equiv \tilde{M}$}
\label{fig:kappaindi}
\end{figure}

\begin{figure}
\begin{center}
\includegraphics[width=10.0cm]{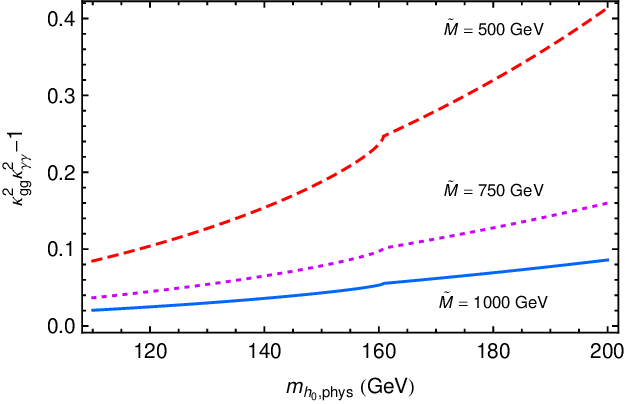}
\end{center}
\caption{\small \em The relative change in the cross-section times decay rate
  for the full process $gg \to h_0 \to \gamma \gamma$ in the LWSM, expressed
  as $|\kappa_{gg}|^2 |\kappa_{\gamma \gamma}|^2 -1$, plotted as a function of
  $\mph{h_0}$.  Lee-Wick mass scales are such that $M_Q = M_u =
  \mph{\tilde{h}} = \mph{\tilde{h}+} = \mph{\tilde{W}} \equiv \tilde{M}$}
\label{fig:kappatot}
\end{figure}

\begin{figure}[t]
\begin{center}
\includegraphics[width=10.0cm]{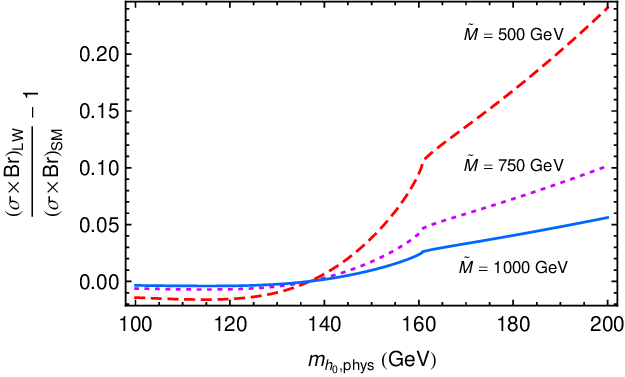}
\end{center}
\caption{\small \em The relative effect on the experimentally measurable
  quantity of cross section times branching ratio $\sigma(gg\to h_0)\,{\rm
    Br}(h_0 \to \gamma \gamma)$ in the LWSM, plotted against $\mph{h_0}$ for
  different values of the LW scale $\tilde M$.}
\label{fig:sbr}
\end{figure}

In Fig.~\ref{fig:rscale} the dimensionless correction factors $s_{H}(r_{h_0})$,
$s_{H-\tilde H}(r_{h_0})$, (defined in Eq. \eqref{eq:s}) and $s_{(A +
  \tilde{A})^2}(r_{W})$ (defined in Eq. \eqref{eq:sA}) are plotted as
functions of $r_{h_0}$ or $r_{W}$, respectively. Notice that as $r\to 1$ the
correction factors increase sharply, corresponding to the case where the LW
and SM masses become degenerate. In this limit, interference effects between
the LW and SM contributions to a process should be carefully treated. For this
work, the masses of the SM and LW Higgs bosons will be assumed to be
sufficiently well separated such that only the production and decay of an
on-shell SM-like Higgs will need to be considered.

Constraints on the various LW gauge boson mass scales were discussed in Ref.
\cite{GOW} and in section \ref{sec:bosons}. In particular, experimental
constraints on $\Delta \rho$ placed bounds on the mass $M_1$ (corresponding to
the mass of the LW heavy ``photon'' in the limit that $M_{(1,2)} \gg
g_{(1,2)}\,v)$ such that $M_1 \stackrel{>}{{}_{\sim}} 1\, {\rm TeV}$. However,
since there is no tree-level constraint from muon decay on the LW $\tilde W$
mass (corresponding to $M_2$ in the limit $M_{(1,2)} \gg g_{(1,2)}\,v$) we
will assume $M_2$ could be lower than 1~TeV. Similar assumptions will also be
made for the LW top quarks and the LW Higgs bosons such that the analysis
can be performed with a common LW scale defined as, $\tilde{M} \equiv M_Q =
M_u = \mph{\tilde{h}_0} = \mph{\tilde{h}+} = \mph{\tilde{W}}$.

The upper plot in Fig.\ \ref{fig:kappaindi} shows the relative change in the
cross-section, with respect to the SM, for the processes $gg \to h_0$ in the
LWSM. This is displayed as a function of the SM-like Higgs boson mass
$\mph{h_0}$. The lower plot in Fig. \ref{fig:kappaindi} shows the relative
change in the decay rate $h_0 \to \gamma \gamma$ in the LWSM, also as a
function of the SM-like Higgs boson mass $\mph{h_0}$. It is immediately
apparent that the effects on the cross-section $gg\to h_0$ are much larger
than those on the rate $h_0 \to \gamma \gamma$.  This can be traced back to
the dependence of $\kappa_{gg}$ on $s_{H-\tilde H}$, which rises strongly when
the SM Higgs mass $\mph{h_0}$ approaches the LW Higgs mass
$\mph{\tilde{h}_0}$.

For Higgs boson masses $\mph{h_0} =$ 130~GeV the correction to the rate for
$gg \to h_0$ varies between $+4\%$ and $+16\%$ for LW masses $\tilde{M}$
between 1000~GeV and 500~GeV, respectively.  For Higgs boson masses even
closer to the LW scale the corrections can be much larger, cf.\ Fig.\
\ref{fig:kappaindi}.

The corrections to the decay rate $h_0 \to \gamma \gamma$ are smaller but
negative for the same Higgs mass $\mph{h_0} = 130$~GeV. For LW masses
$\tilde{M}$ between 1000~GeV and 500~GeV the corrections are between $-0.8\%$
and $-2.8\%$, respectively.  Notice that these corrections do not rise
strongly for larger Higgs masses (in the range [110 -- 200]~GeV) because the
dominant contribution to the process $h_0 \to \gamma \gamma$ comes from
$W$-boson loops which have a coupling to the Higgs boson modified by the much
more slowly rising $s_H$ correction factor.

Fig.\ \ref{fig:kappatot} shows the relative change in the cross section times
decay rate, $\sigma (gg\to h_0)\,\Gamma(h_0 \to \gamma \gamma)$, Eq.
\eqref{eq:sigmagamma}. For $\mph{h_0} = 130$~GeV with a LW mass scale
$\tilde{M} = 500$~GeV, the enhancement in the rate is $\sim12\%$.  For
$\mph{h_0}$ closer to the LW scale the rate is enhanced even more due mainly
to the enhancement of $gg \to h_0$ discussed above.

However, the quantity with direct impact on experimental measurments is more
likely to be the total rate
\begin{equation}
\sigma(gg \to h_0) {\rm Br}(h_0 \to \gamma\gamma)\,.
\end{equation}
The relative change in this quantity is plotted in Fig.\ \ref{fig:sbr}.  The
effects are found to be smaller than the changes in the cross section times
decay rate because of the simultaneous enhancement of the total Higgs width in
the LWSM. This enhancement almost cancels the effects of the larger cross
section times decay rate for lighter Higgs masses and is responsible for the
overall small reduction in the total rate for Higgs masses up to about
135~GeV.

The Higgs width in the SM $\Gamma(h_0 \to all)$ is well approximated by
\begin{equation}
\Gamma(h_0 \to all) = \Gamma(h_0 \to ( \bar f  f, ZZ,  W^+W^-))\,,
\end{equation}
where $\bar f f$ can be $\bar b b,\,\bar \tau \tau$ or $\bar c c \ldots$ and
potentially off-shell gauge bosons are also included. The total Higgs boson
width in the LWSM can be obtained by scaling the individual SM decay rates
(for the tree level processes). Using Eqs. \eqref{eq:s} and \eqref{eq:sA}, the
vector boson modes are scaled by a factor $s_H^2\,s_{(A+\tilde A) }^2$ and the
$\bar f f$ modes are scaled by a factor $s_{H-\tilde H}^2$.

A technique for measuring the Higgs boson couplings $g\,g\,h_0$ and
$\gamma\,\gamma\,h_0$ at the LHC was discussed in Ref. \cite{georg}. With
2$\times$300 fb$^{-1}$ of integrated luminosity, it was estimated that a new
contribution to the $g\,g\,h_0$ coupling could be measured if the
corresponding partial width was larger than $\pm(30 - 45) \%$ of the SM
expected value, for Higgs boson masses in the range $110 \stackrel{<}{{}_\sim}
\mph{h_0} \stackrel{<}{{}_\sim} 190$~GeV. The prospects for measuring a new
contribution to the $\gamma\,\gamma\,h_0$ coupling are slightly better. If a
new contribution to the partial width for $h_0 \to \gamma \gamma$ is larger
than $\pm(15 - 20)\%$ of the SM expected value, then it should be measurable
for Higgs masses in the range $120 \stackrel{<}{{}_\sim} \mph{h_0}
\stackrel{<}{{}_\sim} 140$~GeV.

Notice from Fig.~\ref{fig:kappaindi} that the expected changes in the $h_0 \to
g g$ and $h_0 \to \gamma \gamma$ partial widths in the LWSM fall just short of
the expected reach of the LHC with 2$\times$300 fb$^{-1}$ of integrated
luminosity. It could be hoped that more luminosity would uncover these small
deviations, however, the apparent absence of a deviation from the SM in these
channels, combined with the presence of one or more heavy LW resonances (such
as a $\tilde W$ \cite{Rizzo:2007ae}) could be taken as evidence for the LWSM.
A further potential signature of the LWSM will be discussed in the next section.

\section{Enhancement of the CKM elements $|V_{\rm t(d,s,b)}|$}
\label{sec:flavour}

As has already been discussed in Sec.\ \ref{sec:Yukawa}, each of the SM
fermions experiences a mixing with the corresponding two chiral LW fermions,
proportional to the ratio
 \begin{equation}
 \varepsilon_{\rm ph} = \phys{m}{f}/ M \,.
 \end{equation}
While this mixing is negligible for most fermions, it may have an important
effect on the top quarks and their LW counterparts, which will affect typical
flavour physics observables sensitive to new physics such as $B\to
X_{d,s}\gamma$
or $B_{d,s}$-mixing.

To see how this works out, consider the interaction of the top quarks and
counterparts in the down-type sector, mediated by $W$ bosons.  The
interaction term in the Lagrangian has the form
\bea
\lefteqn{{\cal L}_{dWt} =
\sum\limits_{i=d,s,b}\left[
        V^*_{it}\bar\Psi^i_L\,\fslash{W} \,\Psi^t_L +
        \mbox{\rm h.c.}\right]}\nnb\\
&=&
\sum\limits_{i=d,s,b}\left[
        V^*_{it}\left(\bar D^i_L,\,\bar{\tilde D}^i_L,\,
                        \bar{\tilde d}^i_L\right)\,
        \fslash{W} \,
        \left(\baccc 1+\frac12\varepsilon_{\rm ph}^2 & \cdots & \cdots\\
                     \vdots & \ddots & \vdots \\
                     \vdots & \cdots & \cdots   \ea\right)
        \left(\bac T_L\\\tilde T_L\\\tilde t_L\ea\right) +
        \mbox{\rm h.c.}\right]
\,.
\eea
This implies that the charged current interactions of any down-type quark with
the top quark, i.e.\ any interaction of the form $V_{it}^{*\rm CKM}\bar
d^i_L\fslash{W}t_L$, in LWSM is enhanced by the same amount
$1+\varepsilon_{\rm ph}^2/2$ w.r.t.\ the original SM value.  Clearly the CKM
matrix is then not by itself unitary anymore.  Unitarity, which is tightly
connected with the GIM mechanism and renormalizability, is obtained when both
LW states and the SM particles are properly accounted for.

Clearly, for the non-top CKM matrix elements, which are obtained from direct
measurements, LW mixing effects are negligible due to the small SM quark
masses.

Up to now, the values of $|V_{td}|$ and $|V_{ts}|$ have only been measured in
rare processes, with the top quark running in loops. Therefore, so far, these
CKM matrix elements have mainly been measured in products with $|V_{tb}|$,
and values for $|V_{td}|$ and $|V_{ts}|$ have been obtained by assuming
$|V_{\rm tb}| = 1$ on the grounds of the unitarity of the CKM matrix. Also,
many of those measurements in fact provide values for $|V_{td}/V_{ts}|$, where
this extra factor due to the diagonalization in the top sector of the LWSM
cancels out exactly.

It is worth noting here that overshooting unitarity is a clear an indication
of the ghost-like nature of the additional degrees of freedom.  In other
models including extra positive-norm particles, such as models with four or
more quark generations and vector-like quarks, unitarity would be undershot,
c.f. Ref.~\cite{Vtb1} for example.

An additional point should be stressed here, which was already hinted at when
commenting on the unitarity of the model above. The rescaling
$1+\varepsilon_{\rm ph}^2/2$ will of course receive corrections from the
propagation of intermediate LW partners. This will lead to a decrease in the
effect. Such mutual cancellations of mixing and additional particles will
probably render constraints on the LWSM from flavour physics less stringent
than naively expected from either the negative contribution of the additional
particles or the LW mixing factors alone.

The situation is different for the measurement of $|V_{\rm tb}| = 1.3(2)$ by
the \DO collaboration at the Tevatron \cite{D0}. The experimental set-up makes
sure that the intermediate vector boson has the mass of the SM $W$ boson and
is not an exotic heavy degree of freedom.  Therefore there will be no
decreasing correction to the scaling factor $1+\varepsilon_{\rm ph}^2/2$.

An overview of constraints on $|V_{\rm tb}|$ and the implication of new
physics models can be found in Ref.~\cite{Vtb1}.

\section{Consequences for collider searches}

The LWSM would manifest itself through a plethora of resonance-like
structures.  It should be stressed again that these resonances are effective
degrees of freedom of a more fundamental theory rather than true particles, as
indicated through their ghost-like nature.

Concerning the Higgs boson spectrum, the LWSM provides a second Higgs doublet
with four additional scalars, like many other models beyond the Standard
Model, consisting of one heavy CP-even and one CP-odd state plus two charged
ones.  The latter three are, at tree-level, degenerate in mass, the CP-even
state is lifted away from the other objects through mixing with the light SM
Higgs boson.

The trilinear vertex of the LW Higgs boson with gauge bosons is enabled
through mixing with the SM boson only.  The relevant mixing factor is
$(s_H-s_{H-\tilde H})$, with $s_H$ and $s_{H-\tilde H}$ defined in
Eq.~\eqref{eq:s}.  In other words, the gaugophobic nature of the LW Higgs
boson depends on the level of degeneracy of the two neutral CP-even Higgs
bosons. So, as a rough estimate, the Higgs sector of the LWSM looks like the
Higgs sector of the MSSM close to the decoupling limit. Hence, in most cases,
the heavy Higgs bosons could be found only though their decay into the
fermions.

Signatures of additional twelve LW quarks could be found in future colliders.
If in addition $M_Q = M_u$ or $M_Q=M_d$, as we have assumed for the top
sector, then the LW fermions would be pairwise mass degenerate.  This
degeneracy in fact is only lifted in the top quark sector, through the
non-vanishing SM top quark mass.  A similar phenomenon would occur for the LW
leptons and neutrinos, but mixing effects would definitely not play a role.

So, at first sight, the situation would look quite similar to the case of the
first level of fermions in models with universal extra dimensions (UEDs)
\cite{Appelquist:2000nn}.  A more careful discussion of this issue, however,
is clearly beyond the scope of the present paper and should be the subject of
forthcoming studies.

In addition to that, the LWSM gives rise to resonances of all gauge bosons,
which could be searched for through standard methods employed in the search of
$Z'$, $W'$ and $g'$ states.  This has already been discussed in
\cite{Rizzo:2007ae} \footnote{ There, however, the relative phase of the LW
  partners of the gauge bosons, which is fixed in the LWSM, has been left open.}.  Finding them would make the similarity to UED models more manifest,
the absence of neutralinos, charginos and gluinos and a mismatch of numbers of
states in turn would help ruling out the MSSM or similar supersymmetric
models.  Of course, again, this statement is to be taken with the caveat of
the particles being kinematically accessible.

In contrast to the two other models mentioned in this section, the LWSM has a
number of distinguishing features: \bit
\item First of all, while both the MSSM and the UED models provide
        a parity ($R$-parity or $KK$-parity), forcing the existence
        of a lightest, stable supersymmetric or KK partner particle,
        all LW partners may decay, such that ultimately only SM-like
        particles emerge.  Hence, typically there won't be any
        signatures - apart from those involving neutrinos in the
        decay chains - involving large missing $E_T$ fractions.
\item Secondly, all LW partners have exactly the same spin as their
        SM counterparts.  A careful measurement of the spins of
        decaying resonance-like structures through momentum correlations
        in their decay chains in the spirit, e.g., of \cite{Barr:2004ze}
        will therefore help ruling out the MSSM.
      \item In order to distinguish the LW model from models like the UED
        model, a careful analysis of the mass spectrum or the particle mixings
        may help.  The truly unique feature of the model is that the ghost
        nature of the LW states leads to cancellations whereas the
        orthosymplectic diagonalization gives rise to an enhancement which
        results in small effects despite the many resonances, as in the
        amplitude $gg \to h_0 \to \gamma\gamma$.  The UED model
        \cite{Petriello} is distinct in that the additional KK states lead to
        purely additional effects.  \eit




\section{Conclusions}
In this paper the process $gg \to h_0 \to \gamma\gamma$ was investigated in the
LWSM, an extension of the SM proposed by Grinstein, O'Connell and Wise, based
on an idea of Lee and Wick. All relevant sectors of the model were
diagonalized by orthosymplectic transformations, differing from the usual
unitary diagonalization procedure. This difference was necessitated by the
ghost-like nature of the LW partners to the SM fields.

Numerically, the changes to the measurable overall cross-section times branching ratio
$\sigma(gg \to h_0)\,{\rm Br}(h_0 \to \gamma\gamma)$ are rather small, due to
compensating effects between the mixing-enhanced couplings and cancellations
due to the ghost-like nature of the LW partners running in loops. It has been
demonstrated how this effect can be understood from the higher derivative
formulation of the theory. In channels other than $h_0 \to \gamma \gamma$ the
larger enhancement of the cross-section $\sigma(gg \to h_0)$ may be
measurable.

Signals of heavy LW resonances at future colliders, taken in combination with
the absence of a significant change in $\sigma(gg \to h_0)\,{\rm Br}(h_0 \to
\gamma\gamma)$ is a distinctive feature of the LWSM.

Further signatures of the LWSM may be found in the top sector, where it has
been shown that measurements of $|V_{t(b,s,d)}|$ would show an enhancement
from the expected value, contrary to the predictions of, for example, 4th
generation models. This enhancement is entirely due to the ghost-like nature
of the LW-particles and constitutes another particularly distinctive
signature.

Besides experimental constraints on the LW gauge bosons, coming from
measurements of $\Delta \rho$, it would certainly be desirable to work out in
more detail constraints on the LW top, W's and Higgs bosons.

\paragraph{Addendum} This version is  
equivalent to the PRD version including the erratum. 
The corrections in Eq.~\eqref{eq:kappaph} do not change the conclusions of the arXiv v2 of this paper.
It is worth noting that
among the three standard electroweak reference parameters $\alpha$, $G_F$ and
$m_Z$, only $m_Z$ receives corrections at tree level \cite{Underwood:2008cr}.
This has been consistently used throughout this work and earlier versions.  It
is in this respect that we disagree with the relevant results in \cite{v2,
  Cacciapaglia:2009ky}. It seems to us that in the arXiv version \cite{v2} the
corrections to $\alpha$ and $G_F$ have not been taken into account in the
range of models considered\footnote{We cannot make any statement of
  whether they are small or not in any other models than the
  LWSM.}. Ironically, since the Lee-Wick model has no corrections to those
parameters at tree level we do agree with the results in \cite{v2} when
expanding our results to leading order in one over the Lee-Wick masses squared. However,
we do not agree with the JHEP version of the paper \cite{Cacciapaglia:2009ky}
where corrections to the Higgs VEV, or equivalently $G_F$, have been
attempted.  More precisely we agree with (3.43) in \cite{v2} but not with
(3.43) in \cite{Cacciapaglia:2009ky}.

\section*{Acknowledgments}

The authors are grateful to Oliver Brein, Stefan F\"orste, Nigel Glover,
Michael Schmidt and Georg Weiglein for discussions. 
We are also grateful to Jose Zurita and Giacomo Cacciapaglia for discussion concerning their work.  In particular, Terrance
Figy was of great help during the beginning stages of this project and
provided some insight into the mixing of the LW counterparts of the
electroweak gauge bosons.  FK thanks the Galileo Galilei Institue for
Theoretical Physics for the hospitality and the INFN for partial support
during the completion of this work.  FK wishes to thank the Marie Curie
research training network {\sc MCnet} (contract number MRTN-CT-2006-035606)
for financial support.  RZ is supported in part by the Marie Curie research
training networks contract Nos.\ MRTN-CT-2006-035482, {\sc Flavianet}, and
MRTN-CT-2006-035505, {\sc Heptools}.  

\appendix
\setcounter{equation}{0}
\renewcommand{\theequation}{A.\arabic{equation}}

\section{Trace property of the orthosymplectic invariant $I_t$}
\label{app:invariant}

In this appendix the unity of the orthosymplectic invariant 
$I_t = {\rm Tr}[ \,  g_t \, \eta_3 \, {\cal M}_t^{-1}\,] $, as stated in 
Eq.~\eqref{eq:It}, will be proven.  It is instructive to use a slightly more 
general form of the matrices $g_t$ Eq.~\eqref{eq:lorentz3} and ${\cal M}_t \eta_3$
Eq.~\eqref{topmass} which will be denoted by a bar.  Also, the subscript $t$ 
for the top will be omitted.
\begin{equation}
\overline{{\cal M}}
\ =\ \left( \begin{array}{ccc}
A_{11} & A_{12} & 0 \\
A_{21} & A_{22} & -M_u \\
0         & -M_Q   & Y \end{array} \right) \quad
\bar g \ =\ \left( \begin{array}{ccc}
A_{11} & A_{12} & 0 \\
A_{21} & A_{22} & 0 \\
0         &  0   & Y \end{array} \right) \,.
\end{equation}
Here, a general $2 \times 2$ matrix $A$ has been allowed in the first two 
indices and the additional Yukawa type interaction Eq.~\eqref{eq:Yukawa_ex}.

Now, the question is under which conditions the modified invariant,
\begin{equation}
\label{eq:condi}
\bar I =  {\rm Tr}[ \, \bar  g  \, \eta_3 \,  \overline {\cal M}^{-1}\,] =
1  \,,
\end{equation}
remains unity.  To answer this, the matrix $\overline {\cal M}$ is rewritten as
\begin{equation}
\label{eq:MQu}
\overline {\cal M} \, \eta_3 =  \bar g -  M_{Qu} \qquad   M_{Qu} \equiv
\left( \begin{array}{ccc}
0 & 0 & 0 \\
0 & 0 & M_u \\
0 & M_Q & 0 \end{array} \right) \,.
\end{equation}
By writing
\begin{equation}
\label{eq:main}
3 = {\rm Tr}[ (  \bar g -  M_{Qu}) (  \bar g -  M_{Qu} )^{-1}] = \bar I  -
 {\rm Tr}[ (  M_{Qu})(  \bar g -  M_{Qu} )^{-1}] \,
\end{equation}
the task then reduces to find the conditions for the last term on the
right hand side to be $-2$. The latter is easily evaluated
\begin{eqnarray}
{\rm Tr} [  M_{Qu}\eta_3  \overline {\cal M}^{-1}  ] =
2 A_{11} M_Q M_u  \det( \overline {\cal M} \, \eta_3 )^{-1} \,,
\end{eqnarray}
observing that only two entries of the ${\cal M}^{-1}$ enter the trace.
The determinant is
\begin{equation}
 \det(\overline {\cal M} \,\eta_3) = - A_{11} M_Q M_u + Y {\rm det}_2(A)
\end{equation}
and therefore Eq.~\eqref{eq:condi} is fulfilled if and only if the determinant of 
the $2 \times 2$ matrix $A$ is zero.  This is the case for the matrix ${\cal M}_t$
Eq.~\eqref{topmass} even when an extra Yukawa term of the type 
Eq.~\eqref{eq:Yukawa_ex} is considered.

\section{Triangle graph function}
\label{app:triangle}

In this appendix the well known results for the triangle graphs for the
subprocesses $h_0~\to~gg(\gamma\gamma)$ as shown in
Fig.~\ref{fig:gluonphoton}, will be given.  All results will be parametrised
in terms of an effective Lagrangian
\begin{equation}
{\cal L}^{\rm eff} =  \frac{- g_2}{16 \pi \, \mph{W}} h_0
\left\{
\begin{array}{l}
\frac{1}{2} \alpha_s G^2 F_{gg}  \\
\phantom{x} \alpha F^2 F_{\gamma\gamma}
\end{array}
\right\}  \,.
\end{equation}
A term $1/2$ has been factored out in front of the gluon field field
strength tensor $G^2 \equiv  G_{\mu\nu}^a G^{\mu\nu \, a}$, which originates
from the normalisation of the Gell-Mann matrices
${\rm Tr}[\frac{\lambda^a}{2} \frac{\lambda^b}{2}] = \frac{1}{2}\delta^{ab}$.
The gluon only couples to the quarks whereas the photon also couples
to $W$-bosons.  The number of graphs contributing to the latter can
be considerably simplified by employing the background field gauge
\cite{background} or a non-linear $R_\xi$ gauge \cite{nonlinearR} by choosing 
the gauge parameter such that Higgs-gauge terms and gauge fixing terms cancel.
The gluon and and photon form factors are parametrised as follows
\begin{equation}
F_{gg} = F_{1/2}  \qquad F_{\gamma\gamma} =  F_{1/2}  + F_1 
\,,
\end{equation}
Ghosts and unphysical charged scalars contributions  have to be added to the $W$ boson contribution
\begin{equation}
F_1 = F_1^W + F_0^\eta +  F_0^\varphi   \, ,
\end{equation}
where $\eta$ denotes the ghosts and $\varphi$ 
the would-be Goldstone bosons.  However, employing the 
non-linear $R_\xi$ gauge \cite{nonlinearR} the individual terms read
\begin{eqnarray}
\label{eq:F}
F_{1/2} &=& - 2 \beta_f[1+ (1-\beta_f)f(\beta_f) ]  \nonumber \\
F_{1}^W & =&  4 \beta_W[1+ (2-\beta_W)f(\beta_W) ]  \nonumber \\
F_{0}^\eta &=&  -\beta_W[1- \beta_W f(\beta_W) ]  \nonumber \\
F_{0}^\varphi & =&  -\frac{2}{\beta_W} F_{0}^\eta
\end{eqnarray}
with threshold parameters $\beta_{x}  = 4\mph{x}^2/\mph{H}^2$ and
\begin{equation}
f(x) = \left\{ \begin{array}{ll}
{\rm Arcsin}^2(1/\sqrt{x}) & \quad x\geq 1  \\[0.2cm]
-\frac{1}{4}(\ln\big(\frac{1+\sqrt{1-x}}{1-\sqrt{1-x}}\big)-i \pi)^2 & \quad x < 1
\end{array} \right. \,.
\end{equation}
The result for $F_{1/2}$ can for instance be found in reference \cite{background}.
For the fermion loops it is only the top quark that matters, the size of the $b$ 
quark loop is below one percent as compared to the top contribution.
The sum of the three contributions of the vector
\begin{equation}
F_1 =  F_1^W + F_0^\eta + F_0^\varphi  =
3 \beta_W(2-\beta_W)f(\beta_W) + 3\beta_W + 2
\end{equation}
reproduces the result in reference \cite{background}.

In the table below values for the form factors are quoted, which are of relevance
for the analysis in this publication, namely the asymptotic values for an infinitely 
heavy loop particle  $\beta\to \infty$, which were first obtained in \cite{higgsold} 
and shown to be the photonic $\beta$-function coefficients \cite{background}, and 
the values for physical threshold parameters for a Higgs mass
$\mph{h_0} = 120\,{\rm GeV}$.
\begin{equation}
\renewcommand{\arraystretch}{1.5}
\addtolength{\arraycolsep}{9pt}
\begin{array}{l | r | r || r | r | r | }
                     & F_{1/2} & F_1 & F_1^W & F_0^\eta &  F_0^\varphi \\
   \hline
 \beta \to \infty &  -\frac{4}{3} & 7 &  \frac{20}{3} & \frac{1}{3} & 0    \\
 \beta_{\rm phys}  \sim   &  -1.37   & 8.2 & 8.3 &  0.5 & -0.6
 \end{array}
\renewcommand{\arraystretch}{1}
\addtolength{\arraycolsep}{-9pt}
\end{equation}

\section{Top quark masses and Yukawa couplings}
\label{app:matrices}
The 3 physical top quark masses squared, $\mph{t}^2$, $\mph{\tilde t}^2$
and
$\mph{\tilde T}^2$ can be conveniently be expressed as the 3 roots of
the
following cubic polynomial
\begin{equation}
x^3 \ -\ \left(M_Q^2 + M_u^2\,\right)\,x^2 \ +\
M^2_Q\,M_u^2\,x \ -\ m_t^2\,M_Q^2\,M_u^2 \ =\ 0\,.
\end{equation}
Most conveniently for phenomenological analyses the top quark Yukawa
couplings
$\phys{g}{t}$ and the physical top quark masses $\phys{{\cal M}}{t}$ can
be
expressed in terms of the physical (SM) top quark mass $\mph{t}$, where
$\varepsilon_{\rm ph}= \mph{t}/M$ and $M_{Q} = M_{u} \equiv M$,
\footnotesize
\begin{eqnarray}
\frac{\phys{g}{t}}{M} &=&
\left(
\begin{array}{ccc}
  \varepsilon _{\rm ph}+2 \varepsilon _{\rm ph}^3 &
  \frac{1}{\sqrt{2}} \varepsilon _{\rm ph} +\frac{3}{4 \sqrt{2}}
\varepsilon_{\rm ph}^2 +\frac{55}{32
    \sqrt{2}} \varepsilon _{\rm ph}^3 &
  -\frac{i}{\sqrt{2}} \varepsilon_{\rm ph}+\frac{3 i}{4
    \sqrt{2}} \varepsilon _{\rm ph}^2 -\frac{55 i}{32 \sqrt{2}}
\varepsilon _{\rm ph}^3\\
  \frac{1}{\sqrt{2}} \varepsilon _{\rm ph} +\frac{3}{4 \sqrt{2}}
\varepsilon
  _{\rm ph}^2 +\frac{55}{32 \sqrt{2}} \varepsilon _{\rm ph}^3 &
  \frac{1}{2} \varepsilon _{\rm ph} +\frac{3}{4} \varepsilon _{\rm ph}^2
+
  \varepsilon _{\rm ph}^3 & -\frac{i}{2}\varepsilon _{\rm ph} -\frac{7
    i}{16} \varepsilon _{\rm ph}^3 \\
  -\frac{i}{\sqrt{2}} \varepsilon _{\rm ph} +\frac{3 i}{4 \sqrt{2}}
  \varepsilon _{\rm ph}^2 -\frac{55 i}{32 \sqrt{2}} \varepsilon _{\rm
    ph}^3 & -\frac{i}{2} \varepsilon _{\rm ph} -\frac{7 i}{16}
  \varepsilon_{\rm ph}^3 &
 -\frac{1}{2} \varepsilon_{\rm ph} +\frac{3}{4} \varepsilon _{\rm ph}^2
- \varepsilon _{\rm ph}^3
\end{array} 
\right)\nonumber\\
&& \ +\ {\cal O}\left(\varepsilon_{\rm ph}^4\right)\,,
\end{eqnarray}
\normalsize
and
\begin{equation}
\frac{\phys{{\cal M}}{t}\,\eta_3}{M} =
\left(
\begin{array}{ccc}
\varepsilon_{\rm ph} & 0 & 0 \\
0 & -1 + \frac {1}{2} \varepsilon_{\rm ph}  + \frac {3}{8}
\varepsilon_{\rm ph}^2 & 0 \\
0 & 0 & -1 - \frac {1}{2} \varepsilon_{\rm ph}  + \frac {3}{8}
\varepsilon_{\rm ph}^2
\end{array}
\right) \ +\ {\cal O}\left(\varepsilon_{\rm ph}^4\right)\,.
\end{equation}

\end{document}